\documentclass[namedreferences]{SolarPhysics}
\usepackage[optionalrh]{spr-sola-addons}
\usepackage{graphicx}        
\usepackage{color}           
\usepackage{url}             

\begin{document}

\begin{article}

\begin{opening}

\title{3D MHD Flux Emergence Experiments: Idealized models and coronal interactions}

\author{A.~W.~\surname{Hood}$^{1}$\sep 
       	V.~\surname{Archontis}$^{1}$\sep
			D.~\surname{MacTaggart}$^{2}$
}
\runningauthor{Hood et al.}
\runningtitle{Flux emergence and coronal interactions}

   \institute{$^{1}$School of Mathematics and Statistics, University of St Andrews, St Andrews, Fife, KY16 9SS, U.K.
                     email: \url{alan@mcs.st-and.ac.uk}\\ 
   $^{2}$ Niels Bohr International Academy, Blegdamsvej 17, 2100 Copenhagen, Denmark\\             
}

\begin{abstract}
{This paper reviews some of the many 3D numerical experiments of the emergence of magnetic fields
from the solar interior and the subsequent interaction with the pre-existing coronal magnetic field. The models
described here are idealized, in the sense that the internal energy equation only involves the adiabatic,
Ohmic and viscous shock heating terms. However, provided the main aim is to investigate the dynamical evolution, this is adequate. Many interesting observational phenomena are explained by these models in a self-consistent manner. }
\end{abstract}
\end{opening}

\section{Introduction}

The rapid increase in computing power over the last decade has meant that it is now possible to run numerical experiments that can provide important information about the physical processes involved in explaining how magnetic fields rise through the final layers of the solar interior and emerge into the solar atmospheres of the photosphere, chromosphere and, ultimately, the corona. Despite the new parallel computing resources, it is still difficult to simulate anything more than the uppermost 10 Mm of the convection zone and perhaps 50 Mm into the corona in the vertical direction and 50 - 60 Mm in both horizontal directions. Nonetheless, over this physical vertical range, the plasma density still varies by over eight orders of magnitude. In addition, the timescales change from the convection timescale of the order of a few minutes to coronal timescales of the order of seconds. Correctly resolving such disparate timescales is computationally challenging. Finally, the forces responsible for controlling the plasma dynamics vary from the pressure and gravity forces in the high plasma $\beta$ interior to the magnetic forces in the low $\beta$ corona. These forces control flux emergence when they are of similar magnitude, namely when the plasma beta, $\beta = 2\mu p/B^2$, is of order unity in the photosphere.

There are various physical processes that are important in the different regions of the Sun, depending on the values of the local plasma properties, such as density and temperature. The convection zone involves a highly turbulent plasma that generates the granules and supergranules, with lifetimes of the order of 5 min and 12 h respectively. In the photosphere, the temperature is sufficiently low that the plasma is only weakly ionized. In addition, the transport of thermal energy requires detailed radiative transfer processes to be included. While this is important for the thermodynamics of the plasma and for direct comparison of plasma properties with observations, the impact is not so important for the dynamics of the magnetic field. Anisotropic thermal conduction and optically thin radiation are important in the corona. Indeed the assumption of a collisional plasma becomes less true the higher into the corona the experiments extend. All of these processes can be modelled numerically but the disparate timescales makes this a daunting task. Instead, this paper will review the substantial progress that has been made by restricting attention to the magnetic field evolution, how it emerges and how it interacts with the pre-existing coronal magnetic field. Once fully emerged from the solar interior, it is the magnetic field that controls the dynamics of the plasma in the solar corona.

Much of the pioneering work on simulating flux emergence was started by Shibata and co-workers
(\citeauthor{shibata89a} \citeyear{shibata89}, \citeyear{shibata89a}, \citeyear{shibata90}, \opencite{shibata92}, \opencite{yokoyama95}). 
While this work was mainly 2D, it laid 
the foundations for the subsequent 3D simulations. \citeauthor{matsumoto1998} \shortcite{matsumoto1998} were the first to 
investigate the emergence of a 3D flux tube from the the solar
interior. They investigated how a cylindrical flux tube, unstable to a kink instability \cite{hood79}, can rise from
the interior into the solar atmposhere. For a previous summary of flux emergence in 2D and 3D
see \inlinecite{archontis08b}. However, there has been significant progress in understanding flux emergence over 
the last couple of years, bringing many different ideas together. The new observations from \textit{Hinode} and 
theoretical advances have been discussed at a series of flux emergence workshops, with the most recent results 
summarised in this issue. It is time to review the theoretical aspects of flux emergence again.

The paper has the following outline: In Section \ref{SEC:1}, the basic equations, used in all simulations of magnetic field emergence, are stated. A short description of the standard choices of initial states and the typical boundary conditions used in the simulations are discussed. Section \ref{SEC:SUBPHOTO} describes the initial evolution of buoyancy initiated flux emergence. Section \ref{SEC:Photosphere} presents the typical features and phenomena that occur in simulations and observations at the photosphere and Section \ref{SEC:AXIS} discusses the interesting problem of the evolution of the flux tube axis. A common feature seen in all flux emergence experiments is the formation of a new flux rope above the photosphere in discussed in Section \ref{SEC:ROPE}. The common coronal phenomena of sigmoids and plasma eruptions are described in Sections  \ref{SEC:SIGMOIDS} and  \ref{SEC:ERUPTIONS}. The conclusions are presented in Section \ref{conclusion}.


\section{Basic Equations}\label{SEC:1}
The usual resistive MHD equations for a plasma in the solar corona are used:
\begin{eqnarray}
\rho{\partial {\bf v} \over \partial t} + \rho \left ({\bf v}\cdot \nabla \right){\bf v} &=& - \nabla p +{\bf j} \times {\bf B} + \rho {\bf{g}}  \label{1} \;\;,\\
{\partial {\bf B} \over \partial t} &=& \nabla \times \left ({\bf v} \times {\bf B} \right) + \eta \nabla^{2} {\bf B}\;\;,\label{2}\\
{\partial {\rho} \over \partial t} + \nabla \cdot \left (\rho{\bf v} \right) &=& 0\;\;,\label{3}\\
{\partial p \over \partial t} + {\bf v} \cdot \nabla p &=& - \gamma p \nabla. {\bf v} + (\gamma -1) \frac{j^{2}}{\sigma}\;\;,\label{3b}\\
\mu \:{\bf{j}} &=& \nabla \times {\bf{B}}\;\; ,\label{4}\\
p &=& \frac{1}{\tilde{\mu}}\rho R T\;\; , \label{5}
\end{eqnarray}
where ${\bf v}$  is the plasma velocity, ${\rho}$  is the mass density, ${p}$  is the gas pressure, ${\bf B}$  is the magnetic induction (usually called the magnetic field), ${\bf{j}}$ is the electric current density, $T$ the temperature, ${\bf{g}}$ is the gravitational acceleration, ${\gamma}$  is the ratio of specified heats, ${\eta}$  is the magnetic diffusivity and is related to ${\sigma}$ the electrical conductivity and $\mu$ the magnetic permeability through $\eta = 1/\mu \sigma$, $\tilde{\mu}$ is the mean molecular weight ($\tilde{\mu} = 0.5$ for a fully ionised hydrogen plasma, $\tilde{\mu} = 1$ for a neutral hydrogen gas), $R$ is the gas constant. Different authors have used various codes to solve these. Although the codes use different algorithms to solve the MHD equations, they all produce consistent results. Some codes ignore resistivity, since it is \textit{not possible} to model the physically relevant value and just use numerical diffusion to allow magnetic reconnection to occur. Others prefer to control the physical processes by using an enhanced value of resistivity. This ensures that reconnection occurs through a controllable physical process rather than numerical effects. Steep gradients and shocks inevitably occur during the simulations and codes must be sufficiently robust to handle them. For example, the LARE code \cite{arber01} uses shock viscosities to resolve shocks and the associated shock heating is included in the internal energy equation, Equation (\ref{3b}). 

\subsection{Basic Equilibrium}\label{mageqn}
The basic initial state is assumed to be in equilibrium  and so it must satisfy
\begin{equation}
\nabla p = {\bf j} \times {\bf B} + \rho {\bf g}.
\label{eq:equilib1}
\end{equation}
Since the gas pressure appears linearly in Equation (\ref{eq:equilib1}), it is common practice to split the pressure into a background component that balances the gravity term and a second component that balances the Lorentz force. The background plasma is determined first by prescribing the temperature, as discussed below.
\subsubsection{Background Plasma without a Magnetic Field}
The majority of simulations assume that the background plasma is determined by imposing the temperature, $T(z)$, as a function of height. Then, using the gas law Equation (\ref{5}), the equation of hydrostatic balance,
\begin{displaymath}
\frac{dp_{\rm{b}}}{dz} = - \rho_{\rm{b}} g,
\end{displaymath}
can be integrated to give the background pressure, $p_{\rm{b}}$, and background density $\rho_{\rm{b}}$. It is typical to assume that the temperature profile is linear in the solar interior, with a temperature gradient that is either equal to or just greater than the adiabatic value required for the onset of convection. The photosphere and corona are taken as uniform, with a chromosphere and sharp transition region in between. Considering a background plasma all the way from the upper region of the convection zone right up to the corona, means that the simulation codes have to deal with density variations of the order of 8 orders of magnitude.
\subsubsection{Initial Sub-Photospheric Magnetic Field}
A magnetic field is included next in the solar interior. While it is possible to include a force-free magnetic field, this is complicated and instead the magnetic forces are balanced by an additional pressure force. This additional modification to the gas pressure, $p_m$, satisfies
\begin{equation}
\nabla p_{m} = \frac{1}{\mu}\left (  \nabla \times {\bf B}\right )  \times {\bf B}.
\end{equation}
Therefore, Equation (\ref{eq:equilib1}) is satisfied if $p = p_{\rm{b}} + p_{\rm{m}}$. Normally $p_{\rm{m}}$ is negative so that this term must not be too large, in order that the gas pressure remains positive. In a high $\beta$ plasma, as is appropriate in the solar interior, this is not a problem. However, including an ambient coronal magnetic field in a low $\beta$ plasma must be done with care in order to avoid such problems with the gas pressure. Different forms for the initial magnetic field are discussed below.

The plasma is now in force balance but, since $p_m$ is normally negative, the temperature (proportional to $(p_{\rm{b}} + p_{\rm{m}})/\rho_{\rm{b}}$) within the magnetic field is now lower than the background value. If isotropic thermal conduction in the solar interior is significantly large, then the plasma would very soon equalize  the temperature inside and outside the magnetic regions and, hence, for the given gas pressure, $p_{\rm{b}} + p_{\rm{m}}$, the density must be lower than the surrounding background density. In this situation, the magnetic region becomes buoyant. 

What form does the interior magnetic field take? It cannot be observed at present and so simple models are chosen to initiate flux emergence in different contexts. These are now discussed.
\begin{enumerate}
\item To model the \textit{small-scale} emergence in granules and supergranules, the magnetic field is not so well organized and it is sufficient to take an initial, horizontal magnetic sheet in the solar interior of the form ${\bf B} = ( 0, B(z), 0)$. Most researchers assume a layer of finite thickness. The sheet is then disturbed either by adjusting the density, as described above, or by specifying an initial velocity distribution to produce many small-scale emergence regions. An example of this procedure is described in \inlinecite{archontis09a}.

\item To model \textit{active regions}, it is usual to assume that the magnetic field is generated initially at the base of the convection zone in the tachocline. Magnetic buoyancy transports a flux tube from there to the top of the convection zone. In order for the magnetic field to remain coherent during this rise, the flux tube must be twisted by a sufficient amount \cite{moreno96,emonet98}. Hence, many researchers \cite{fan01,magara03,manchester04,archontis04,murray06,dmac10} have placed a horizontal, \textit{cylindrical, twisted flux tube} near the top of the convection zone. In cylindrical coordinates, $( r, \theta, y)$, where $r^2 = x^2 + (z - z_{\rm{axis}})^2$ and $z_{\rm{axis}}$ is the height of the flux tube axis, the initial magnetic field has the form
\begin{displaymath}
{\bf B} = \left ( 0, B_\theta (r), B_y(r)\right ),
\end{displaymath}
and the axis of the loop lies along the horizontal $y$ axis. The standard cylindrical flux tube used in the majority of simulations is that of \inlinecite{fan01}, namely
\begin{displaymath}
B_y = B_0 e^{-r^2/d^2}, \quad B_\theta = \alpha r B_y.
\end{displaymath}
This field has a constant twist and, since it is not force-free, there is a perturbation to the pressure of the form 
\begin{displaymath}
p_m = B_0^2 e^{-2r^2/d^2} (\alpha^2 d^2 - 2 - 2\alpha^2 r^2)/4\mu. 
\end{displaymath}
$B_0$ is the field strength on the axis of the tube and $d$ is a measure of the radius of the flux tube. The pressure perturbation is always negative if $\alpha d < \sqrt{2}$.

Then the flux tube is made buoyant, as above, to form an $\Omega$-shaped loop that rises to form a large bipole. Typically a Gaussian profile is used for the density deficit, $\rho_{\rm{m}}$, of the form
\begin{equation}
\frac{\rho_{\rm{m}}}{\rho_{\rm{b}}} = \frac{p_{\rm{m}}}{p_{\rm{b}}} e^{-y^2/\lambda^2},
\label{dens_deficit}
\end{equation}
where $y$ is the distance along the axis of the tube and $\lambda$ measures the length of the buoyant section of the tube. However, it is clear that the whole of the tube is weakly buoyant, even for $y > \lambda$. Note that $p_m \rightarrow 0$ as $r \rightarrow \infty$ so that there is no pressure deficit at the edge of the tube and the total gas pressure merges into the background value. In addition, $p_m$ is proportional to $B_0^2$ and so there is no buoyant plasma if the magnetic tube is removed.

\item To model the emergence of the top part of a $\Omega$-shaped loop that is rooted much deeper in the interior, an alternative choice of initial flux tube is a toroidal shaped loop. \inlinecite{hood09} and \inlinecite{dmac09a} describe how to construct an initial toroidal loop. Consider a toroidal flux tube, with the major radius of the torus, $R_0$, larger than the minor radius, $d$. The form of the magnetic field to leading order in an expansion in powers of $d/R_0$ is based on the cylindrical model. Thus,  we use first a cylindrical coordinate system $( R, \phi, -x)$, where $R^2 = x^2 + (z - z_{base})^2$ and $z_{base}$ is the base of the computational domain, and then a local toroidal coordinate system $(r, \theta, \phi)$, where $r^2 = x^2 + (R- R_0)^2$. The toroidal magnetic field components are $B_\phi(r)= B_0 e^{-r^2/d^2}$ and $B_\theta(r) = \alpha r B_\phi$, \textit{i.e.} the same as the cylindrical case above. Finally, the cartesian components of the magnetic field are, to $O(d/R_0)$,
\begin{eqnarray}
\label{torusx}
B_x &=& B_\theta (r)\frac{R - R_0}{r}, \\
\label{torusy}
B_y &=& - B_\phi \frac{z - z_{base}}{R}  + B_R \frac{y}{R},\\
\label{torusz}
B_z &=& B_\phi \frac{y}{R} + B_R\frac{z - z_{base}}{R},
\end{eqnarray}
where $B_R = - B_\theta(r) x/r$. One important consequence of using a toroidal loop is that the field near the axis does not possess a dip that would collect dense plasma. Hence, any dense plasma that subsequently forms is free to drain down the field to the base of the computational box.
\end{enumerate}

\subsubsection{Initial Coronal Field}
One key aim of flux emergence simulations is to understand how the emerging magnetic field interacts with the overlying coronal magnetic field. Except for isolated null points, there is nowhere in the solar corona, where the magnetic field is negligible. There are two simple possibilities. One is to assume that the coronal magnetic field is uniform and horizontal, vertical or slanted and the other is a non-uniform field. 

A simple model for the overlying coronal field is to assume that the field is horizontal and of the form
\begin{displaymath}
{\bf B} = B_0(z) \left (\cos \theta, \sin \theta, 0\right ),
\end{displaymath}
where the field strength, $B_0(z)$, increases from zero at the top of the photosphere to a constant coronal value at the base of the corona. Choosing different values for $\theta$ means that the overlying field and the emerging field can interact at any angle between parallel and anti-parallel. This is discussed in detail in \inlinecite{archontis05} and  \citeauthor{galsgaard05} (\citeyear{galsgaard05}, \citeyear{galsgaard07}).

To include a non-uniform coronal field in the initial equilibrium, there are two basic approaches.  The first is to insert an analytical magnetic field.  Since the model for the atmosphere (solar interior to corona) is highly stratified, this makes the calculation of (suitable) full magnetohydrostatic equilibria challenging.  
The simplest choice is to use a potential field.  In many idealized models of flux emergence, the solar interior is free of 
any ambient magnetic field.  This is chosen to isolate the dynamics of the emergence of the tube at the photosphere.  If 
a potential field is included in the initial condition, it will fill the entire domain.  However, if the potential field has a 
simple enough geometry and the flux function is known analytically, magnetic field lines can be removed from the 
solar interior in the initial equilibrium.  Then, the field will need to relax to a nearby equilibrium. As an example, consider the simple potential arcade defined by
\begin{displaymath}
\mathbf{B} = \nabla\times(A\hat{\mathbf{y}}), \quad A = B_0 l e^{-z/l}\cos(x/l),
\end{displaymath}
where $A$ is the flux function, $B_0$ and $l$ are parameters and the magnetic field is invariant in the $y$-direction.  To make a field-free region in the solar interior, field lines beginning (and ending) in some horizontal region, $[-x_1,x_1]$ say, on the bottom boundary, must be removed.  To determine the points $(x,z)$ that lie in this region, we trace down field lines to the lower boundary.  Using the fact that the flux function is constant on field lines, a field line passing through the point $(x,z)$ will touch the lower boundary at
\begin{displaymath}
x_0 = l\cos^{-1}\left(e^{(z_{base}-z)/l}\cos(x/l)\right),
\end{displaymath}
where $z_{base}$ is the base of the computational domain.  If $-x_1 < x_0 < x_1$, then the magnetic field at $(x,z)$ can be removed.  Repeating this test for all $(x,z)$, produces a field-free region in the solar interior.  With the removal of magnetic field from the solar interior, a skin current forms between the field-free zone and the magnetic field.  This field will not be in exact equilibrium but can be allowed the relax to a nearby equilibrium. Since the interior is a high $\beta$ plasma, this should only involve a small change to the field. A flux tube can then be inserted and rise to the photosphere unimpeded.  This technique can be extended to more complex fields, such as laminated equilibria \cite{low82}.

The second approach is to use a numerical simulation to produce an initial coronal field. \inlinecite{archontis08a} consider a model where a flux tube emerges and fills the atmosphere with a non-uniform magnetic field.  Then, they emerge another flux tube into the expanded field of the first.  This enables the study of dynamic behavoiur such as current sheet and plasmoid formation.  A similar technique was employed by \inlinecite{dmac10} in order to produce an initial equilibrium arcade for a flux tube to emerge into.  The arcade is formed by emerging two flux tubes, with no $\Omega$-loop enhancements, at different times, \textit{i.e.} one tube is placed lower in the solar interior than the other.  When the first tube emerges, its magnetic field expands into the atmosphere.  Later, when the second tube emerges, it expands into the overlying field of the first.  The twist of the two tubes is chosen to be in the same direction.  This means that upon impact, their respective fields are in opposite directions.  As the emerging field of the second tube pushes into that of the first, a current sheet forms.  Eventually, reconnection takes place and the system relaxes into a quadrapolar equilibrium.  The X-point is high in the corona and beneath it is an arcade anchored in the two flux tubes at the top of the solar interior.  This equilibrium satisfies the conditions of a magnetic arcade anchored in the solar interior and a field-free region beneath it, in the solar interior, where a flux tube can be inserted and rise without interference from an ambient field. 


\subsection{Boundary Conditions}\label{boundaryconditions}
Boundary conditions are notoriously difficult to impose and frequently the choice of boundary conditions influences the long term evolution of the plasma. The majority of emergence simulations assume that the computational box is periodic in all variables in both horizontal directions. 
This is a reasonable assumption and is particularly useful for setting up the initial magnetic field structures. However, it does mean that the boundaries do begin to influence the subsequent evolution when the magnetic field expands significantly in the horizontal direction and begins to reach the boundary. 
This is not an issue in the interior.

The boundary conditions on the bottom boundary are often chosen as rigid wall conditions with no flow there. Normally, the lower boundary is not an issue as the timescales for the dynamical evolution of the magnetic field slow down as the density rises. The top boundary requires a more detailed discussion. As discussed above, the initial magnetic field is made buoyant and this causes an initial impulse as the flux tube immediately responds to the unbalanced forces. This impulse propagates density and velocity disturbances from the interior into the photosphere and up into the higher atmosphere. As they enter the lower density regions, the velocity amplitude starts to rise and a shock forms. This shock continues to propagate into the corona and interacts with the top boundary. Ideally one would like perfect flow through boundary conditions, so that all outward propagating disturbances simply continue on their way. However, it is extremely difficult to do this for both linear disturbances and shocks. Some codes use an open boundary condition, based on assuming that the normal derivative is zero, but this will still cause a partial reflection of waves from the boundary. A full implementation of characteristic boundary conditions would be needed for the MHD system.  This is computationally difficult to do. Instead, boundary conditions that allow either shocks or linear waves (but not both) to propagate through are used. Alternatively, a damping layer can be introduced that removes kinetic energy in a layer near the upper boundary. This greatly reduces the reflection of waves but does not completely eliminate them. Studies of many simulations indicate that the partial reflection of the initial shock, while causing the corona to be continually excited, does not influence the actual emergence of the magnetic field. However, once the emerging magnetic field reaches the upper boundary, any reflection of it does modify the subsequent evolution of the following magnetic field structures. The simulations should be stopped at that stage.

\subsection{Assumptions and Modelling Philosophy}
To use the results of flux emergence to reproduce the detailed observations obtained from various instruments requires a complete modelling of (i) radiative transfer effects in the photosphere and chromosphere and (ii) optically thin radiative losses, thermal conduction and heating in the corona. In addition, in the cool photosphere and low chromosphere partial ionization must be considered. All of these processes require knowledge of the ion abundances, whether the plasma is in local thermodynamic equilibrium or not, whether the plasma is in ionization balance and so on.

However, the dynamical evolution of flux emergence depends on the timescale of the Lorentz, pressure gradient and gravitational acceleration forces in comparison to the timescale for the above physical processes. For example, at typical coronal temperatures and densities, the timescale for thermal conduction is the order of $500$ s, that for optically thin radiation is approximately $3000$ s (see \opencite{hood92}) while the Alfv\'en timescale is much shorter at $10-60$ s. So the interaction of emerging fields with pre-existing coronal fields can be modelled initially without detailed modelling of the thermodynamic processes. Similarly, once the magnetic field is strong enough to trigger the magnetic buoyancy instability at the photosphere, the forces are predominately due to the excess magnetic pressure and the dynamical rise is solely dependent on the Lorentz force. The thermodynamic variables, such as the gas pressure, do not significantly alter the subsequent rise of the magnetic field to the corona.

Finally, controlled numerical experiments allow one to specify which physical effects are to be studied. Hence, the modelling philosophy is to consider a simpler initial state, for example one that is in equilibrium, and then modify it in a controlled manner to see how one effect at a time influences the evolution. Too many variations all at once obscures the important or dominant physical processes.

\section{Sub-Photospheric Behaviour}\label{SEC:SUBPHOTO}
Assuming that the temperature profile in the interior is linear with $T = T_0 (1 - m z/H)$, where $T_0$ is the temperature at the base of the photosphere ($z=0$), $H$ is the photospheric pressure scaleheight and $m/H$ is the temperature gradient, the buoyant magnetic field will continue to rise towards the photosphere provided the background temperature stratification satisfies
\begin{equation}
-\frac{dT}{dz} \ge - \left (\frac{dT}{dz}\right )_{ad} = \frac{\gamma - 1}{\gamma H}.
\label{buoyancy}
\end{equation}
The initial rise towards the photosphere has been described in detail in \inlinecite{archontis04}. They show how the axial field, $B_y$, satisfies the relation $B_y/\rho = \hbox{constant} $ in 3D simulations in the same way as shown in 2D \cite{emonet98}. During this stage, the azimuthal magnetic field increases at the top of the flux tube, resulting in an increase in the pitch of the field lines when they reach the photosphere. Thus, in all buoyant cylindrical flux tube simulations, the magnetic field always emerges at the photosphere in an initially North-South direction before stretching out into an East-West direction. This is a direct consequence of the twist in the flux tube, needed to ensure the tube remains coherent during its rise through the convection zone.

Since the plasma $\beta$ is very large in the interior, it is normal to assume that the buoyancy is not strongly dependent on the magnetic field. However, since the initial density disturbance does depend on the strength of the magnetic field, an investigation of the effect of varying the field strength was undertaken by \inlinecite{murray06}. By rescaling both the velocity of the axis of the flux tube (rescaled in terms of the Alfv\'en speed) and the time (rescaled in terms of an Alfv\'en timescale), they showed that all the different cases lay on the same curve. This self-similar nature of the buoyancy had been noted in 2D by \inlinecite{emonet98}. In addition, \inlinecite{murray06} also showed that the area of the emerging flux at the base of the photosphere, as a function of time, was self-similar, with time rescaled in terms of the Alfv\'en time. Thus, the properties of the emerging field depend entirely on the strength of the magnetic field in the interior.

The constant twist profile, introduced by \inlinecite{hood79} during investigations into the stability of line-tied coronal loops and by \inlinecite{fan01} for flux emergence experiments, is used by the majority of modellers it seems. However, \inlinecite{murray07} investigated the emergence of flux tubes with a variable twist. Two different twist profiles were considered, one with increasing twist as a function of radius and one with decreasing twist. The main effect of varying the twist profile was to adjust the importance of the magnetic tension force. Flux tubes with twist profiles that have high tension remain coherent and emerge at the photosphere in an almost identical manner. There is only a significant difference between the photospheric appearance of the different twist profiles when the tension forces are lower.

\begin{figure}
\centering
\includegraphics[scale=0.3]{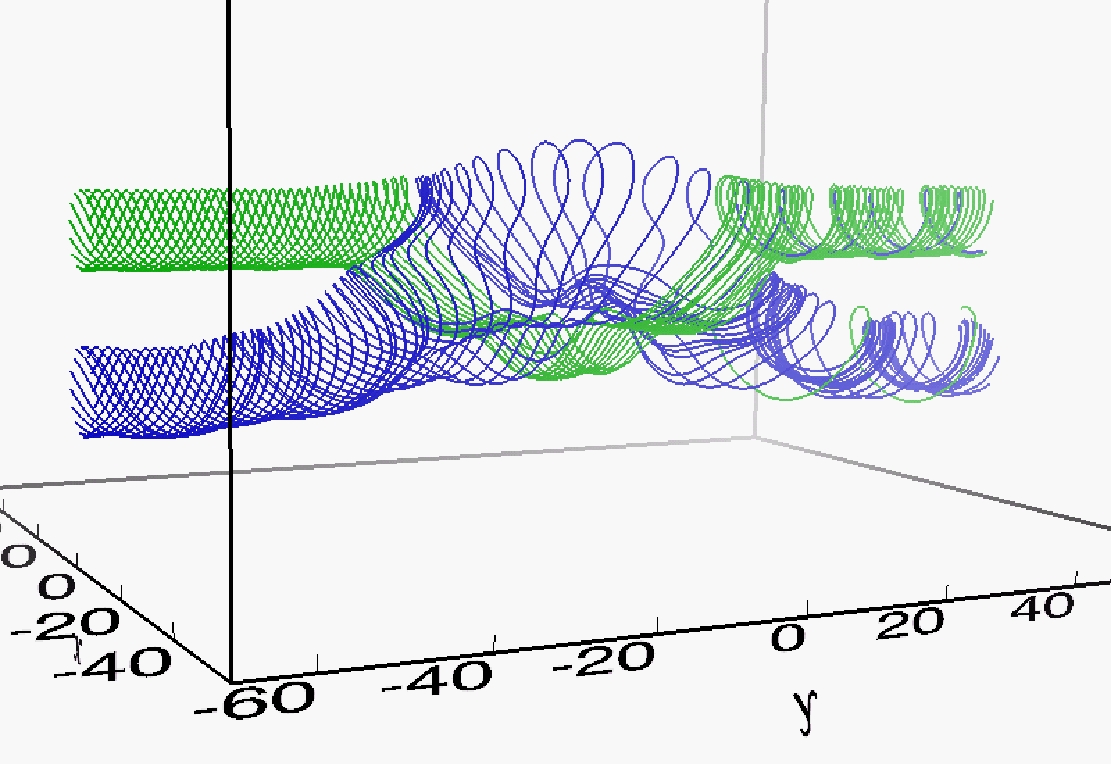}
\caption{Two cylindrical flux tubes. The lower tube is buoyant and has risen towards the photosphere, passing through the upper neutrally buoyant tube. From Murray and Hood (2007).}
\label{complexB}
\end{figure}

Finally, most flux emergence simulations only consider a single flux tube. However, there are many observations showing repeated emergence in a single active region. \inlinecite{murray07} considered two sub-photospheric flux tubes and investigated how a buoyant tube, beneath a flux tube that is in force balance, could emerge through
this magnetic barrier. In fact, the lower buoyant tube tended to simply pass through the non-buoyant tube, as shown in Figure \ref{complexB}, in a manner similar to that described by \inlinecite{linton06} and it was difficult to observe any difference at the photosphere between the various tube orientations. 


\section{Photospheric Phenomena}\label{SEC:Photosphere}
Once the flux tube has reached the base of the photosphere, the subsequent evolution can be compared with a variety of potentially observable features. These are discussed now.
\subsection{Emergence into the corona}
\begin{figure}
\centering
\includegraphics[width=0.49\textwidth]{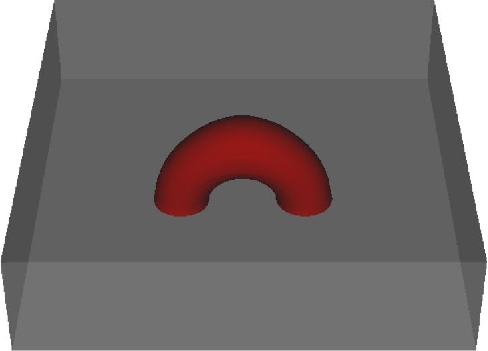}
\includegraphics[width=0.49\textwidth]{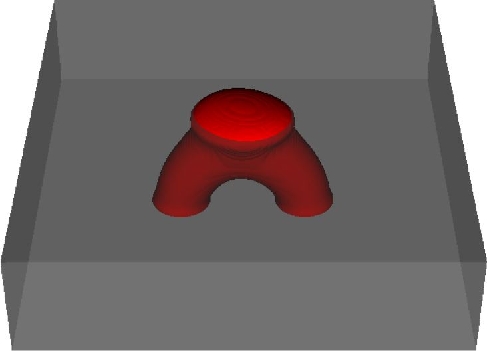}
\includegraphics[width=0.49\textwidth]{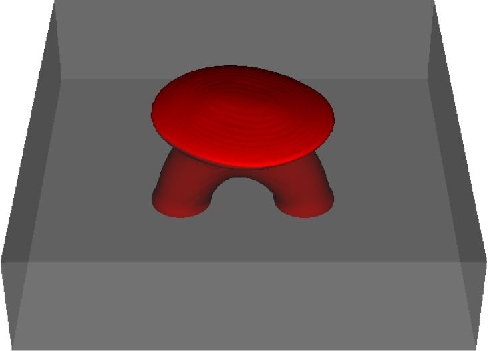}
\includegraphics[width=0.49\textwidth]{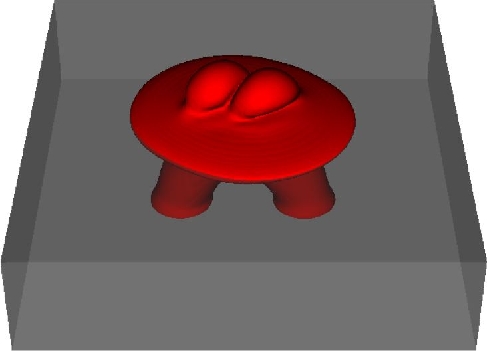}
\label{emerge}
\caption{The contour gives the magnitude of the magnetic field at normalised times 20 (top left), 40 (top right), 60 (bottom left) and 80 (bottom right). The grey region indicates the solar interior out to 1 pressure scaleheight above the base of the photosphere. }
\end{figure}

In the solar interior, the plasma is basically buoyant but, at and above the photosphere, the temperature gradient is no longer decreasing and the plasma is no longer buoyant. Hence, the magnetic field finds itself in an atmosphere where it can no longer continue rising since Equation (\ref{buoyancy}) is no longer true. For the magnetic field to fully emerge and reach up into the corona, a new criterion must be satisfied. The magnetic buoyancy condition is fully discussed in \inlinecite{archontis04} and the instability condition is derived in  \inlinecite{acheson79}. The criterion is
\begin{equation}
-H_p \frac{\partial}{\partial z}(\log B) >  - \frac{\gamma}{2}\beta \delta + \tilde{k}_\parallel^2 \left ( 1 + \frac{\tilde{k}_z^2}{\tilde{k}_\perp^2}\right ),
\label{magbuoyancy}
\end{equation}
with $\delta$ the superadiabatic excess, $\delta = \nabla - \nabla_{ad}$, $\nabla$ the logarithmic
temperature gradient and $\nabla_{ad}$ its adiabatic value, and $\tilde{k}_\parallel$, $\tilde{k}_\perp$
the wavenumbers of the perturbation in the two horizontal directions parallel and perpendicular, respectively, to the magnetic field in units of the local pressure scaleheight. This corrects a typing error in \inlinecite{archontis04}, 
whereby $k_z$ and $k_\perp$ were interchanged. Fortunately, this whole term is not significant. In the photosphere,
$\delta$ is negative and strongly stabilizing. As the magnetic field reaches the photosphere, the left hand side
of Equation \ref{magbuoyancy} (which is positive) involving the logarithmic derivative of $B$ increases but, at the
same time, the plasma $\beta$ reduces. Typically, the onset of the magnetic buoyancy instability occurs when the plasma $\beta$ drops to order unity,  for example see Figure 6 in \inlinecite{fan09}, and the inequality satisfied. 
Only then does the field start to rise from the photosphere and, because the magnetic pressure now exceeds
the background gas pressure, emerge into the corona. This is clearly illustrated in Figure \ref{emerge}, where the contour shows the same value of the magnitude of the magnetic field. At time $t=20$ the magnetic field has yet to reach the photosphere. At $t=40$ the magnetic field has reached into the photosphere by around 2 scaleheights. The field does \textit{not} reach any higher by $t=60$ but instead it starts to spread horizontally. The condition for the onset of the magnetic buoyancy instability has not yet been achieved. However, by $t=80$, the plasma $\beta$ is around unity, the logarithmic derivative is larger and the secondary emergence begins. At this time the field has reached the base of the transition region. Note that in this case, there are two regions that emerge and not just one. What is also clear from this simulation is that there is a significant amount of magnetic flux trapped at the photospheric level. 

Another important consequence of the conditions needed for emergence is that the Lorentz forces are now sufficiently strong that the magnetic field can generate flows at the photospheric level.

\subsection{Sunspot Tails, Separation and Rotation}
\begin{figure}
\centering
\includegraphics[width=0.49\textwidth]{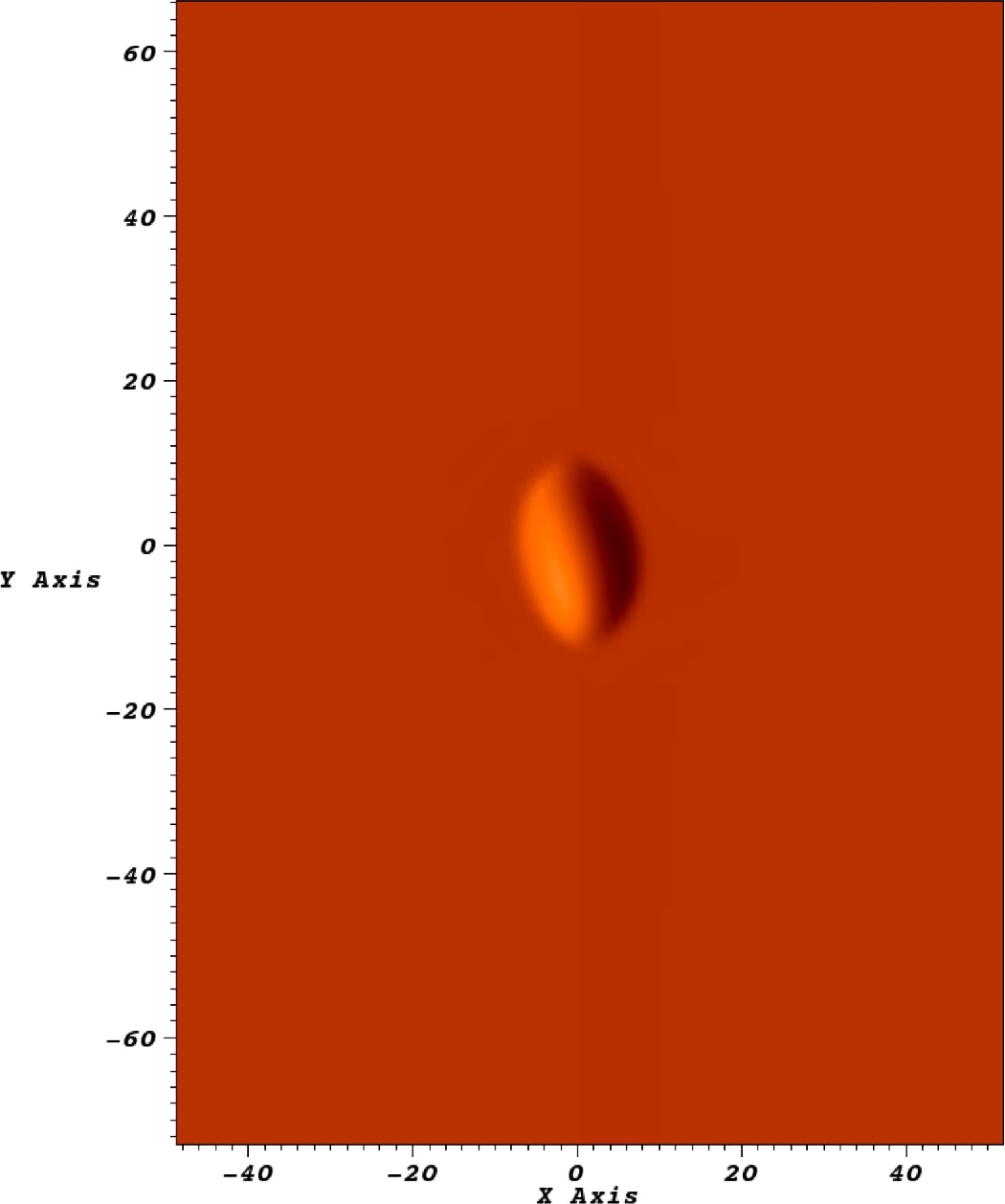}
\includegraphics[width=0.49\textwidth]{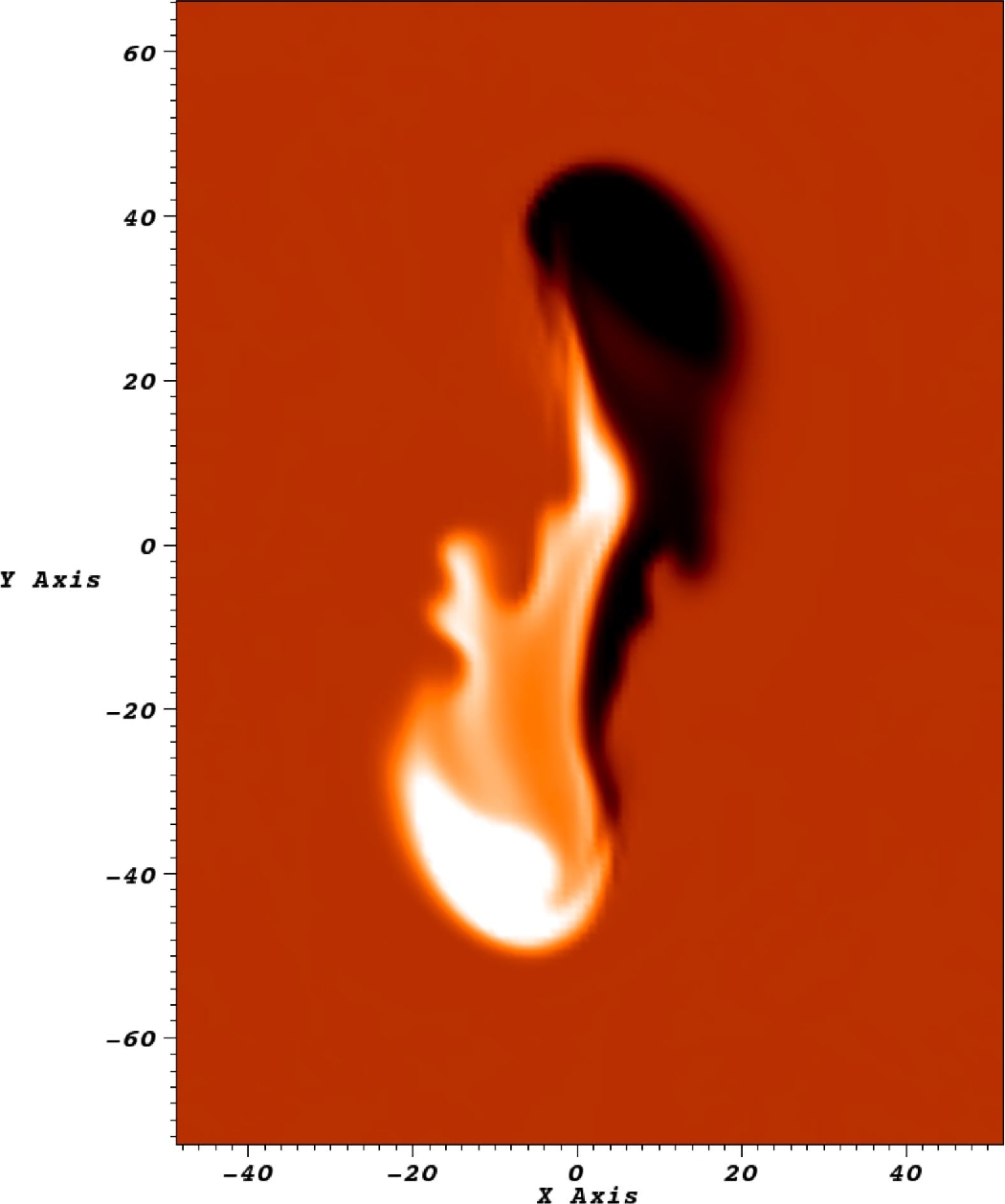}
\label{bztails}
\caption{The line-of-sight magnetogram, showing the initial emergence (left) and
the subsequent development of the magnetic tails (right). From Archontis and Hood (2010).}
\end{figure}

When the field initially emerges at the photosphere, the bipole is
normally oriented in a north-south direction but the dominant opposite
polarities start to drift in an east-west direction. The standard
cylindrical flux tube often shows extended regions trailing from
regions of high magnetic flux (see Figure \ref{bztails}). The formation and dynamical evolution
of magnetic tails in ideal magnetic flux emergence experiments is discussed in \inlinecite{archontis10} and compared with the observations of \inlinecite{canou10}. 
When the flux tube is
highly twisted, the tails are well defined and constitute robust
features of the  emerging flux region. When the twist in the flux tube
is weaker, the tails are less pronounced and easily disturbed by
photospheric flows. Hence, the presence or absence of magnetic tails
in an active region can provide information about the  amount
of twist present in the emerging fields.

Observations of active regions have shown that sunspots emerge and
start to drift apart in opposite directions. The observed separation of the
sunspots only continues for a certain time but an  explanation of when
and why this separation ceases to occur is still missing.  In
numerical experiments that use cylindrical twisted flux tubes, the
magnetic  field is made buoyant by reducing the tube's density through
Equation (\ref{dens_deficit}).  In fact, using this initial condition, the
tubes are buoyant along  their entire length. Hence, in many cases,
the sunspots continue to separate as the emergence process continues
until they reach the horizontal boundaries of the numerical
domain. This is obviously  unrealistic. In the toroidal flux tube
models, on the other hand, the two polarities separate until they
reach the width of the footpoints at the base of the computational
domain.  A similar behaviour can be seen in the cylindrical flux tube
models if the profile of the density reduction in the tubes is adjusted \cite{dmac09b}. What this
suggests is that  the observed final separation of sunspots in new active
regions is giving useful information about the initial buoyancy profile
of the emerging field.

Another feature seen in ideal flux emergence simulations is the
rotation of the sunspots. \inlinecite{magara06} studied the photospheric
motions during the  process of the emergence of a twisted flux tube and
found that, as the  emerging field becomes vertical, a torsional flow
appears in each of the  polarity flux concentrations. It was found
that the polarity region  experienced an apparent rotation opposite to
the torsional flow. This  may indicate that the motion driven in the
polarity region is far from a  rigid rotation. In similar experiments,
\inlinecite{fan09} confirmed that significant rotational/vortical  motion
sets in within each polarity, reminiscent of observations of  sunspot
rotations \cite{brown03,zhao03,tian06,yan08}.  \inlinecite{fan09} also reported that the rotational motions
of the two  polarities are due to torsional Alfv\'en waves  propagating
from the  twisted flux tube into the less stressed corona. So although
the coronal  magnetic field rapidly tries to reach a force-free
equilibrium,  significant energy is transported into the corona
through rotation.  \inlinecite{fan09} found that these motions persist
throughout the later  phase of the evolution and steadily inject
magnetic helicity into the atmosphere.

In the model by \inlinecite{hood09}, the authors found a rotation of
nearly 360 degrees during the emergence of a toroidal flux tube.
The dependence of the rotation on the amount of twist, present in the
emerging field, and the dynamical evolution of the vortical
photospheric  motions for different emerging field configurations are
presently under investigation. A closer comparison between the
aforementioned numerical  experiments with observations of rotating
sunspots should be carried out  in the future.

\subsection{\lq Sea-Serpent\rq\  Emergence and Coronal Response}

High-resolution observations have revealed that the photospheric
distribution  of magnetic flux in emerging flux regions can be
interpreted in terms of  multiple undulations \cite{bernasconi02,georgoulis02,par04,harra10}. 
The \lq sea-serpent\rq\   topology (\textit{i.e.} a series of connected {U}
and $\Omega$ segments of  magnetic field lines) was first suggested
by \inlinecite{harvey73} to explain the  undulating topology of
moving magnetic features near sunspots. The wavelength  of the
undulations of the \lq sea-serpent\rq\  field has been found to be larger
than  2~Mm and less than 10~Mm. This size is larger than  the typical
length scale for  granulation and less than
supergranulation. Therefore, one may assume, naturally, that  the scale
of the undulations is determined by the properties of some magnetic instability.
A suitable candidate is the magnetic buoyancy instability, since its maximum growth rate occurs for wavelengths,
$\lambda$, that are in the range $10H-20H$, where $H$ is the photospheric scale height ($\approx$ 200~km).
This process was shown in the 2.5D numerical
experiments by \inlinecite{isobe07}, who used a  magnetic flux sheet
as the initial configuration for the emerging field. In previous 2D ideal
simulations, \inlinecite{shibata89} have also showed that sub-photospheric
magnetic flux sheets become  unstable to the undular mode of the
Parker instability and form loops that rise into the  corona.

The observations by \inlinecite{par04} suggest that there is a close
connection between the  location of the U-shaped field lines in
undulating systems and the location of Ellerman Bombs (EBs),  which
are small-scale brightenings in the upper photosphere/chromosphere and
brief emissions  in the wings of the $H\alpha$ line. \inlinecite{par04}
suggested that coronal loops may form as  a result of emergence of
undulating field lines that they reconnect at the U-dip locations.  The
brightenings, that are produced at the regions where reconnection
occurs, may account for the observed appearance of Ellerman
Bombs. The above scenario was verified in 2.5D and 3D simulations  by
\inlinecite{isobe07} and \inlinecite{archontis09a}. They find that
magnetic reconnection in the low atmosphere can explain some of the
main characteristics of the EBs, such as the local density and
temperature enhancement. They also report on the formation of long
magnetic systems (similar to long arch filament systems) in the
corona, which are formed by the process of resistive emergence  at
successive atmospheric heights. In the low atmosphere this
process is of fundamental importance.  The magnetic field is free to
emerge and expand, because the dense plasma, which accumulates at the
dips between undulations, is removed from the rising field lines in the low atmosphere by
reconnection. With this mechanism,  the longer reconnected field lines
are no longer anchored to the lower atmosphere and are free to rise into the
corona.  The simulations by \inlinecite{archontis09a} reveal that the
dense plasma, which was initially trapped in the  U-dips, stays below the
photosphere in newly-formed, small twisted flux tubes. In the 2D
simulations by \inlinecite{isobe07}, these structures  look like
magnetic islands that sink rapidly within the unstable
sub-photospheric layer.

The simulations by \inlinecite{archontis09a} also show how the initial undular modes of the Parker instability
develop nonlinearly into interchange-like modes for the  magnetic field and the 
excitation of non-linear dynamic coupling between the
fields emerging from different regions. This produces  a complex network of rising loops that emerge
into the photosphere over a range of temporal and spatial scales. The
interaction of multi-scale loops of emerging flux across the solar
atmosphere can generate a series of  interesting phenomena, for example: i) the
formation and gradual dissipation of currents at various scales and
the formation  of bright features (\textit{e.g.} EBs, bright points, etc) at
different atmospheric heights, ii) the emission of bi-directional
flows that may account for x-ray reconnection jets, iii) the
formation of cool and dense plasma ejections that may  be due to
compression of interacting, pressure-driven, magnetic fields and
correspond to $H\alpha$ surges,  iv) events of flux cancellation at
the low atmosphere and intensification of the magnetic field (both the
vertical and the horizontal components) due to partial evacuation of the
plasma, converging plasma motions and resulting compression of the  magnetic field. 

The interaction of emerging fields has also been studied in the
context of individual flux tubes  that rise separately within the
convection zone and interact above the surface. These
studies have  shown the formation and ejection of plasmoids
\cite{archontis07a,archontis08a}, the triggering of eruptions and associated
solar activity \cite{archontis08a}, the
ejection of high-velocity reconnection  outflows in active regions
\cite{gontikakis09} and the recurrent emission of jets \cite{archontis09b}. More details on eruptions
are presented in Section \ref{SEC:ERUPTIONS}.


\subsection{Adiabatic Expansion}
The energy equation, Equation (\ref{3b}), is essentially an adiabatic equation with ohmic and viscous shock heating. During the emergence process and once the onset of the magnetic buoyancy instability is triggered, there is an excess magnetic pressure that causes a rapid expansion of the plasma. The ohmic and viscous shock heating terms are small and the plasma evolves adiabatically. Hence, the rapid expansion causes the density and pressure to drop (keeping $p/\rho^\gamma$ constant). From the gas law, this results in a substantial drop in the temperature to unrealistically low values. During this stage,  other thermodynamic processes are important. However, there are indications that there is some cooling during emergence, both in simulations with additional 
physics and in observations, but not to the extent predicted by the adiabatic cooling. Nonetheless, the gas pressure, both inside the expanding plasma and in the background atmosphere, remains smaller than the magnetic pressure of the emerging magnetic field and so inaccuracies in the energy equation do not influence the dynamical evolution. Over a longer time, ohmic heating, due to reconnection in forming a new flux rope as discussed below, causes this cool plasma to heat and it eventually reaches a temperature in excess of the background value.

\subsection{Partial Ionization}
The assumption of a fully ionized plasma throughout the whole computational domain is likely to be violated in the photosphere and low chromosphere. Here a substantial fraction of the plasma will be neutral hydrogen. To date the only  3D simulations including the effect of partial ionization are those by \inlinecite{arber07}. The key effect of partial ionization is the anisotropic nature of the resistivity, with collisions between charged particles and neutrals creating an enhanced resistivity perpendicular to the magnetic field. This resistivity, due to Cowling conductivity, results in the rapid dissipation of the perpendicular currents in the photosphere and chromosphere. Hence, by the time the magnetic field reaches the corona there is only a parallel current remaining and the magnetic field is already force-free.

In addition, to removing the perpendicular current, partial ionization also modifies the gas law and, for given pressure and density, increases the temperature, resulting in a larger pressure scaleheight. Thus, the pressure and density do not drop off due to gravity as quickly as for a fully ionized plasma.


\section{The Axis of the Flux Rope}\label{SEC:AXIS}
In simulations of flux rope emergence, the axis of the rope can be clearly identified in the initial condition.  The evolution of the rope axis throughout emergence, however, depends critically on the choice of the initial condition. \textit{i.e.} on the original geometry of the flux rope. First, consider the cylindrical model described in Section \ref{mageqn}, which has been used in the vast majority of 3D flux emergence simulations.  \inlinecite{fan01} considers such an initial flux tube with the imposed buoyancy profile from Equation  (\ref{dens_deficit}).  After emergence, the original axis is found to be constrained to a height of approximately two photospheric pressure scale heights above the base of the photosphere by the end of the simulation.  Using the same setup, \inlinecite{murray06} investigate how changing the initial axial magnetic field strength $B_0$ influences the height the axis can attain.  They find that the larger the value of $B_0$, the higher the height that the original axis reaches.  This is, in part, due to the initial buoyancy force being proportional to $B_0^2$.  Although there is a range of final axis heights dependent on $B_0$, these are still only around two to three photospheric pressure scale heights above the photospheric base.  \inlinecite{magara03} also have a twisted cylinder for their initial condition.  Their setup, however, differs slightly from \inlinecite{fan01} and \inlinecite{murray06}.  They consider a Gold-Hoyle flux rope and impose a velocity perturbation to initiate its rise in the solar interior.  Due to the different internal structure of this flux rope compared to that of Section \ref{mageqn}, they find that the original tube axis expands upwards.  This, however, is still constrained to only reach into the lower atmosphere, i.e. it does not reach coronal heights.  This result appears to be robust for all studies that use a cylindrical flux rope in the initial condition, even when variations are included \cite{arber07,fan09}.

\begin{figure}
\centering
\includegraphics[scale=1]{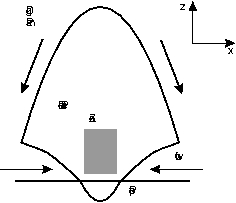}
\caption{A cartoon showing a slice of the emerging magnetic field in the $(x,z)$-plane.  The draining and resulting inflows are indicated.  The reduced pressure region is represented by a grey box. The axis of the original rope lies in the $y$-direction.  Based on MacTaggart and Hood (2009a).}
\label{pressure_zone}
\end{figure}

Another approach to modelling flux rope emergence has been to use a toroidal tube (Equations (\ref{torusx}) - (\ref{torusz})) in the initial condition.  The legs of this tube are anchored at the base of the computational domain and are orthogonal to those of the cylindrical rope. \inlinecite{hood09} and \inlinecite{dmac09a} have investigated the basic dynamics of toroidal flux emergence.  They find that for $B_0$ greater than a threshold value, the original axis of the toroidal tube can emerge to the corona.  For values of $B_0$ below this threshold, the axis remains trapped below the corona, as for the cylindrical flux tube.  
This begs the question: how can the axis of the toroidal flux tube emerge to the corona, when that of the cylindrical model for similar choices of parameter values cannot?  \inlinecite{dmac09b} identify the main cause to lie with the major difference between the two models - their geometry.  When a flux tube emerges, the emerging arcade expands rapidly into the atmosphere due to the steep drop of pressure with height.  This results in a reduction in the total pressure $(p + |\mathbf{B}|^2/(2\mu))$ at the centre of this arcade above the photosphere and is illustrated in Figure \ref{pressure_zone}.  Draining plasma then flows into the region of reduced pressure and can follow one of two paths.  The first is to collect and form dips on field lines in the region of reduced pressure.  The second is to flow down the legs of the flux tube.  For the first path, the original axis of the tube is often inside the region of reduced pressure when plasma drains into it.  This plasma weighs down the axis and contributes to preventing any further rise.  For toroidal tubes with large enough $B_0$, the buoyancy force is strong enough to enable the original axis to reach the top of the reduced pressure zone before draining plasma pins it down. The reason why this does not happen for similarly strong $B_0$ in the cylindrical model is related to the lack of a second path. 

\begin{figure}
\centering
\includegraphics[scale=0.7]{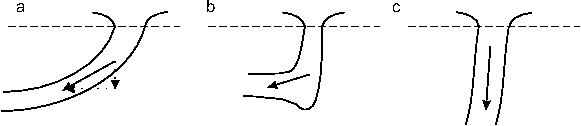}
\caption{The geometry of flux rope legs in the solar interior during emergence.  The base of the photosphere is shown as a dashed line and the beginnings of the expanded fields are shown above it.  (a) and (b) represent the leg geometries for the cylindrical model.  The $\Omega$-loop in (b) is steeper than that in (a) and a dip has formed.  (c) represents the leg geometry of a toroidal loop.  The downward forces of the draining plasma are shown for each case.}
\label{legs}
\end{figure}

Figure \ref{legs} illustrates the geometry of the legs for the cylindrical and toroidal models.  These images convey the shape of the flux tubes in the interior during emergence.  
Figure \ref{legs} (a) represents the leg of an emerging cylindrical rope.  The direction of the downward force of draining plasma is shown as a solid line with its vertical and horizontal components as dotted lines.  With this 
geometry, there is always a vertical force acting to constrain the axis of the rope.   Figure \ref{legs} (b) shows the geometry of another cylindrical tube but with a steeper $\Omega$-loop.  Similar constraining forces are present 
here as for the case in (a).  Due to the sharp turn of the tube (almost a right angle), a dip can form where dense plasma collects. Figure \ref{legs} (c) shows the geometry of a toroidal rope leg.  Here the plasma can drain downwards efficiently and there is little or no component of the force that constrains the axis.   As mentioned 
before, if $B_0$ is large enough, the original axis of the toroidal tube can rise to the corona.  If it is not, the axis is trapped in the lower atmosphere by plasma draining into the reduced pressure zone.  For the cylindrical model, 
the combination of plasma draining on top of the axis and the formation of dips due to the inefficient draining of plasma down the tube, prevents the axis from emerging into the corona.

\begin{figure}
\centering
\includegraphics[scale=0.2]{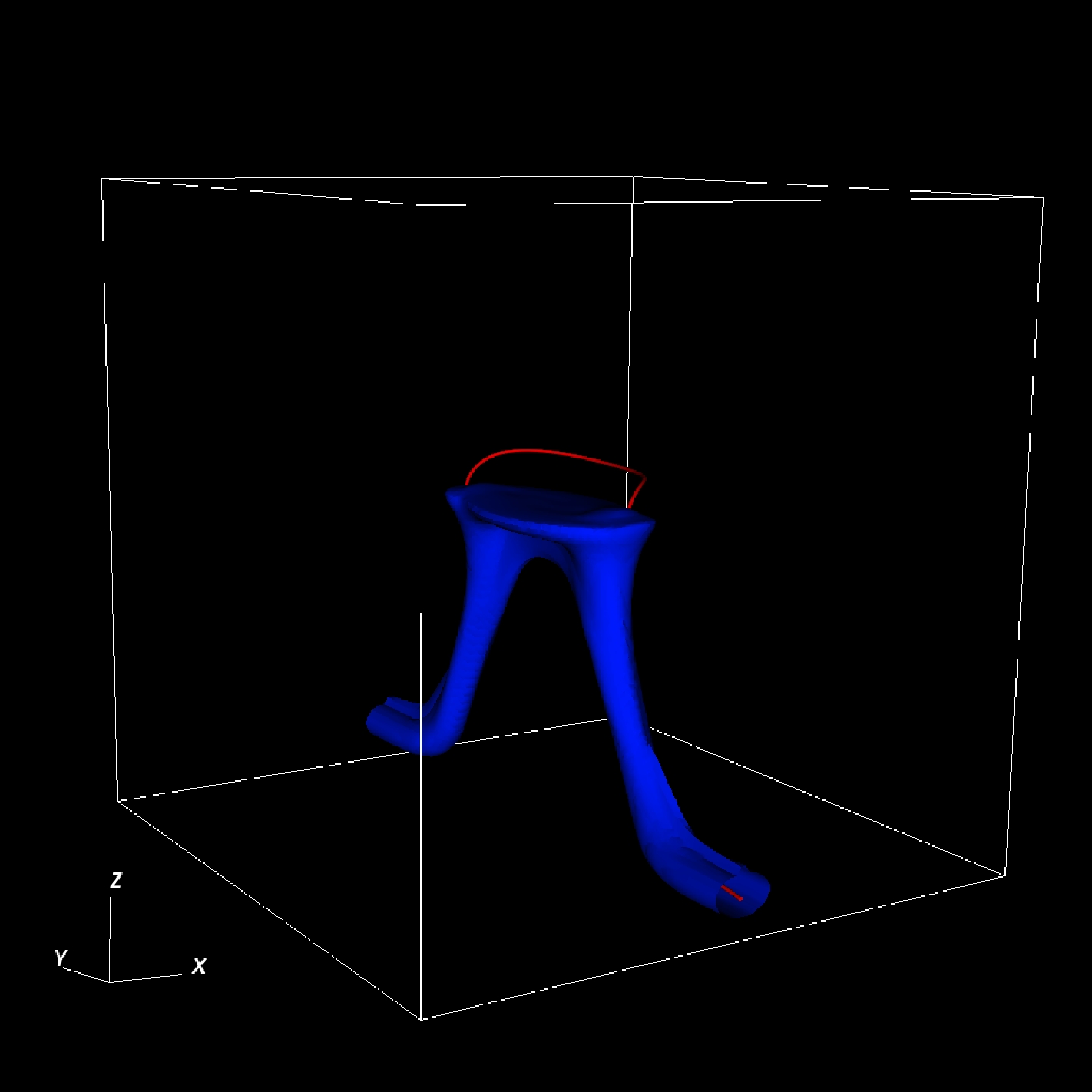}
\caption{The geometry of the cylinder model with the buoyancy profile $n\exp(-y^2/\lambda^2) - (n-1)$ exhibits a toroidal-like shape.  An isosurface of $|\mathbf{B}|=0.5$ in blue and a field line in red indicating the axis.  The flat top of the isosurface indicates the position of the photosphere.  From MacTaggart and Hood (2009a).}
\label{deform}
\end{figure}

One feature of these flux emergence models is that, due to computational constraints, they contain only a small part of the solar interior.  This means that if a cylindrical tube used, it will be placed near the base of the 
photosphere.  With the standard buoyancy profile, a shallow $\Omega$-loop will form, meaning that the legs of the tube will not become steep enough to allow plasma to drain down them efficiently, unlike the toroidal 
model (see Figure \ref{legs}).  One way to modify this is to use a slight variation on the buoyancy profile and to place the cylindrical tube deeper in the interior.  \inlinecite{dmac09a} consider a generalized buoyancy profile of the form

\begin{displaymath}
n\exp(-y^2/\lambda^2) - (n-1),
\end{displaymath}
where $n$ is an integer and the other variables are as explained in Section \ref{SEC:1}.  The effect of this generalized buoyancy profile is to make the central part of the tube buoyant and the ends of the tube over-dense.  If a cylindrical flux tube is placed deep enough in the solar interior, then this buoyancy profile can 
deform it into a toroidal shape.  With this deformation, the geometry of the tube can now allow plasma to drain efficiently down the legs of the tube.  \inlinecite{dmac09a} test this buoyancy profile by placing a cylindrical tube,  
with (using the non-dimensionalization of that work) the parameters $n=6$, $\lambda=20$, $B_0=7$ and $\alpha=0.4$, into a deeper solar interior.  The tube deforms into a toroidal shape with the central part rising and the ends sinking.  Figure \ref{deform} depicts the 
shape of the tube at $t=86$ in the simulation by showing an isosurface of the magnetic field strength, $|\mathbf{B}|=0.5$, and a field line indicating the tube axis.  \inlinecite{murray06} (who use the same non-diensionalization) found that for a tube placed near the photosphere with the 
parameters $B_0=7$ and $\alpha=0.4$, the axis rises to a maximum height of $z\approx 2$.  The top of the axis of the tube in Figure \ref{deform} is at $z\approx 10$.  Hence, it is the geometry of the toroidal model 
that enables the efficient draining of plasma and so allows the axis to emerge further into the atmosphere.   

\section{Flux Rope Formation}\label{SEC:ROPE}
Although, as demonstrated in the previous section, the original axis of the flux tube can emerge into the corona, the geometry of the emerged magnetic field no longer resembles its simple flux rope structure from the initial 
condition.  This, however, does not preclude the formation of new flux ropes above the photosphere.  Indeed, it is the departure of the geometry of the emerged field from that of the initial condition which leads to the formation of new flux ropes in the atmosphere.

As described in Section \ref{SEC:Photosphere}, a flux tube that has risen to the photosphere can emerge into the atmosphere by means of the magnetic buoyancy instability.  The field expands rapidly into the atmosphere due 
to the steep decline of the background gas pressure with height.  As well as a vertical expansion, the magnetic field also expands horizontally.  The rapid expansion of the magnetic field creates a region of reduced pressure (as indicated in Figure \ref
{pressure_zone}) and this acts as a sink for draining plasma.  Apart from this cavitation, the expanding magnetic field produces another dynamically important effect - shearing along the polarity inversion line (PIL) between the 
two main photospheric polarities \cite{manchester01,fan01,manchester04,hood09,dmac09a}.  This shearing can be understood by examining the Lorentz force in the direction ($y$ in this case) of the PIL, namely
\begin{displaymath}
- \frac{\partial}{\partial y}\left(\frac{|\mathbf{B}|^2}{2\mu}\right) + \frac{1}{\mu}\mathbf{B}\cdot\nabla B_y.
\end{displaymath}
The first term is the magnetic pressure force and the second term is the magnetic tension force.  In this flux emergence simulation, it is the tension term that dominates.  In the rapidly expanding field, which is no longer cylindrically symmetric, $B_y$ actually reverses direction.  The gradient of $B_y$ is negative on one side of the arcade and is positive on the other.  Hence, the magnetic tension force drives flows in opposite directions on opposite sides of the PIL.  Note this only occurs when the
Lorentz force is strong enough. However, the field is strong enough since emergence requires the plasma $\beta$ to be of order unity.

With plasma flowing into the region of reduced pressure and shearing occurring along the PIL, reconnection ensues and this leads to the formation of a flux rope.  This process was investigated by \inlinecite{vanb89}, 
where they form a flux rope by \textit{imposing} shearing and cancellation flows on a force-free arcade.  The main difference between their model and the flux emergence models is that, in the latter, the shearing and 
cancellation flows develop self-consistently.  For the cylindrical model, the new flux rope forms above the original axis \cite{manchester04,archontis08}.  For the toroidal model, if the original axis does not exceed the height of the horizontal in-flow, the new flux rope will form above it.  If the original axis does exceed this height, the new flux rope will form below the original axis, interacting and distorting it.  Another feature of the 
toroidal model is that it can produce multiple flux ropes by continuation of the process described above.  If a flux tube emerges into a non-magnetized corona, then any flux ropes that form will not be able to escape (this is discussed further 
in Section \ref{SEC:ERUPTIONS}). In such a model, when a new flux rope is created beneath another rope, the two merge \cite{linton06} and the distinct topologies of the individual ropes are lost.  When an overlying magnetic field is included in the background atmosphere, the first field lines to emerge reconnect with the pre-existing coronal field allowing the newly formed flux ropes to escape.  Again, this will be discussed in Section \ref{SEC:ERUPTIONS}.  

\begin{figure}
\centering
\includegraphics[scale=0.2]{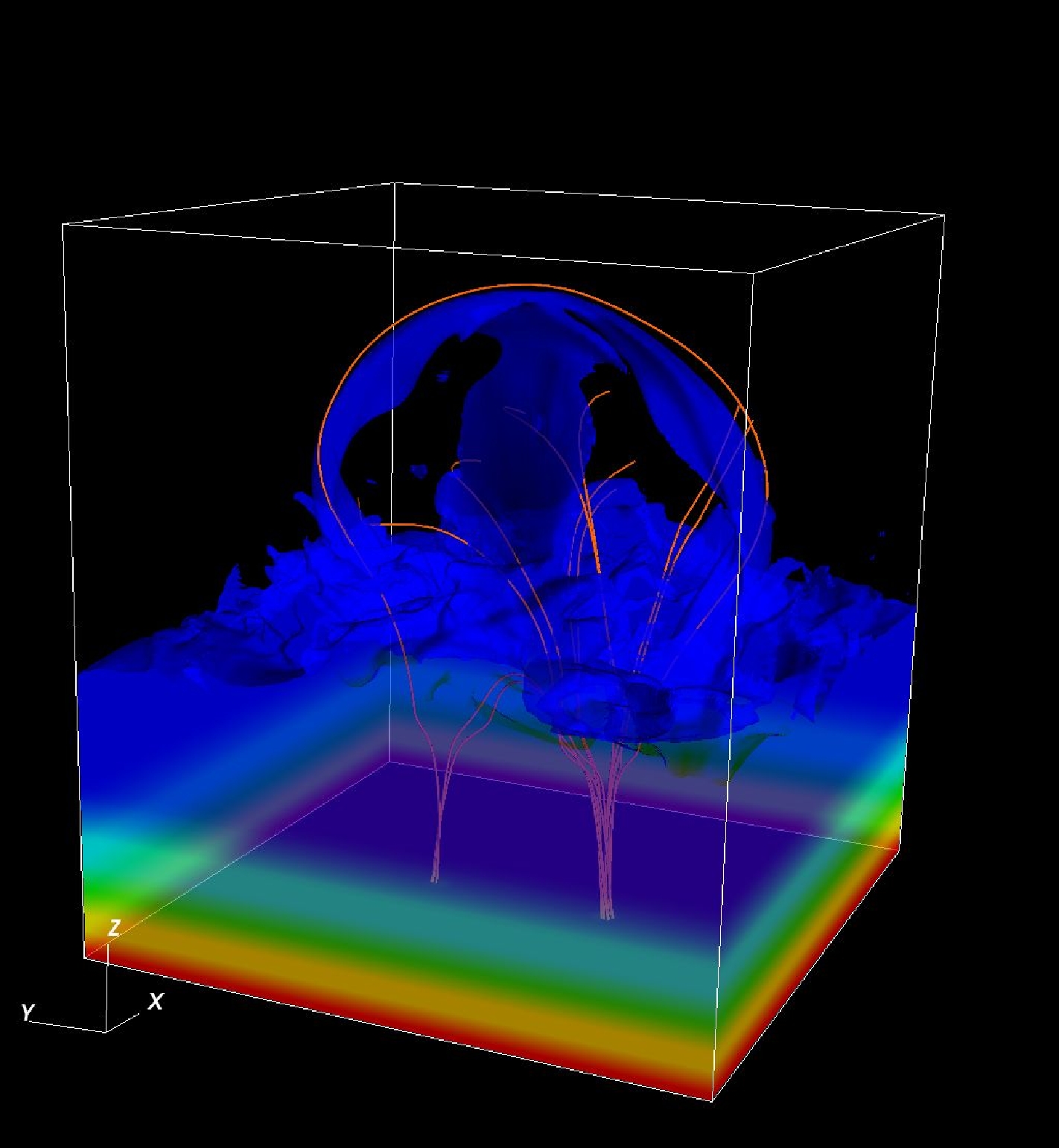}
\caption{A new atmospheric flux rope formed from the emergence of a toroidal rope.  The volume rendering shows density and field lines (in orange) are traced from one of the footpoints of the original toroidal rope.  Based on MacTaggart and Hood (2009b).}
\label{fluxrope}
\end{figure}

Figure \ref{fluxrope} 
illustrates the geometry of a new flux rope.  This is based on \inlinecite{dmac09b}, where an overlying magnetic field is included that does allow any new ropes to escape.  It shows a volume rendering of density.  A loop of dense 
plasma (dense compared to the surrounding coronal plasma) can be clearly seen.  Field lines are traced in orange from one of the footpoints of the toroidal loop.  Some of the field lines connect to the coronal magnetic 
field due to reconnection.  One, however, illustrates how the new rope is weakly twisted, emanating from one side of the new rope and wrapping round behind it on the other side.

\section{Sigmoids}\label{SEC:SIGMOIDS}
\subsection{Introduction}

Ideal experiments of flux emergence, from sub-photospheric layers up into
the corona, have shown that emerging twisted flux tubes  can contain
forward S-shaped and reverse S-shaped field lines
\cite{magara01,fan01,fan03,archontis04}.  For a right-handed
twisted flux tube, the upper part of the windings of the field lines
show an inverse-S shape,  while the lower parts are forward S-shaped,
the latter being consistent with the observations. Thus, one might
expect that sigmoids should be showing the concave upward segments of
the twisted field lines of a flux rope. A common feature in such
simulations is the formation of a sigmoidal current structure, along the 
polarity inversion line, at the dips of sheared and stretched field lines \cite{fan03,manchester04}. \inlinecite{archontis09c} investigated how 
complex sigmoids, which consist of many individual current layers, are 
formed after the emergence of magnetic flux at the solar surface. 

X-ray observations of the solar corona, for example Skylab, \textit{Yohkoh} and \textit{Hinode}, have 
indeed revealed the existence of
structures with  a forward or reverse S-shape. Brightenings associated 
with these structures  were named \textit{sigmoids}
by \inlinecite{rust96}, who also showed that many
of the sigmoidal brightenings evolve into arcades, which are often
associated  with CMEs.
In general, the occurrence of sigmoids in active regions is closely
related to intense solar  activity. Observational studies
\cite{canfield99,canfield07} have  revealed that
active regions with sigmoidal morphology are more likely to lead to
eruptive events (flares or CMEs) than regions that do not possess
sigmoids.

Moreover, observations have shown that many sigmoids have the shape of two {\it
Js} or {\it elbows}, which together form  the forward or reverse
S-shape of the structure. Observational examples of
the different types of sigmoids and reviews on the evolution of
sigmoids can be found in  \inlinecite{canfield99}, \inlinecite{moore01}, \inlinecite{pevtsov02}, \inlinecite{gibson06a} 
and \inlinecite{green07}. 
The complex structure of sigmoids has been reported by 
\inlinecite{mcKenzie08}, who found that sigmoids are not defined by a
single  x-ray loop but instead consist of many loops which together appear
as two {\it J-like} bundles. The later observations have also reported on the  rising
motion of a flux rope-like structure from the middle of the sigmoid
and the x-ray flaring between the two J-shaped systems after the
eruption of the flux rope.

\subsection{Flux Emergence and Sigmoid Formation}

\subsubsection{Shearing and S-Shaped Current }
As discussed in Section \ref{SEC:ROPE}, a shear occurs along the PIL. A result of this shearing is 
that the magnetic field loses its strongly azimuthal nature and begins to run nearly parallel
to the PIL. 

Many simulations \cite{manchester04,gibson06,archontis08} have shown that 
as the magnetic field  rises above the photosphere and expands,
the sheared magnetic field lines are also stretched vertically. Thus,
sheared field lines with opposite  directions come closer together and,
as a result, the current density ($|{\bf j}|=|\nabla \times {\bf B}|$)
becomes large in the region between them.  The current structure, and
the field lines that surround the current, have an   S-like shape
adopting the twist and writhe of the underlying field.
Moreover, the current density initially forms two 
oppositely curved {\it elbows} or {\it J-like} structures. The straight part of the {\it elbows},
undergo shearing and reconnect when they come into direct contact. 
This reconnection of the field lines along the S-like currents produces 
heating, with the hot plasma outlining the sigmoid.

\subsubsection{Field-Line Topology and \lq Bald\rq \  Patches}
Visualization of field lines during the early phase of the emergence, shows 
interesting topological properties. In the models by \inlinecite{gibson06} and \inlinecite{archontis09c}, 
a substantial number of field lines have a concave upward shape at the sites 
where they touch but do not significantly cross the base of the photosphere. 
These sites are called Bald-Patches (BPs). It was found that the magnetic 
topology, which was formed by these field lines is similar to the topological 
structure produced in the model by \inlinecite{titov99}.  The initial
configuration in their model consists of a force-free flux tube with
an arch-like shape that rises quasi-statically into an external
potential magnetic field. Eventually, the tube becomes unstable,
leading to an eruption of magnetic flux.  \inlinecite{titov99} describe
how separatrix surfaces are formed by field lines, which start at a
BP. These surfaces are called  Bald Patch Separatrix Sturfaces (BPSS)
and the field line connectivity has a jump across it. When these
surfaces are  projected onto a horizontal plane, two {\it J-like}
structures are identified, both of them associated with BPs.
In the flux emergence models by \inlinecite{gibson06} and \inlinecite{archontis09c} the 
{\it J-like} bundles of field lines are wrapping around each other along the 
neutral line. They consist of separate sets of field lines but their
projection onto the photospheric plane forms one overall sigmoidal structure.

\subsection {Dynamical Evolution of the Sigmoidal Currents}
\subsubsection{3D Complex Topology}
\begin{figure}
\centering
\includegraphics[scale=0.2]{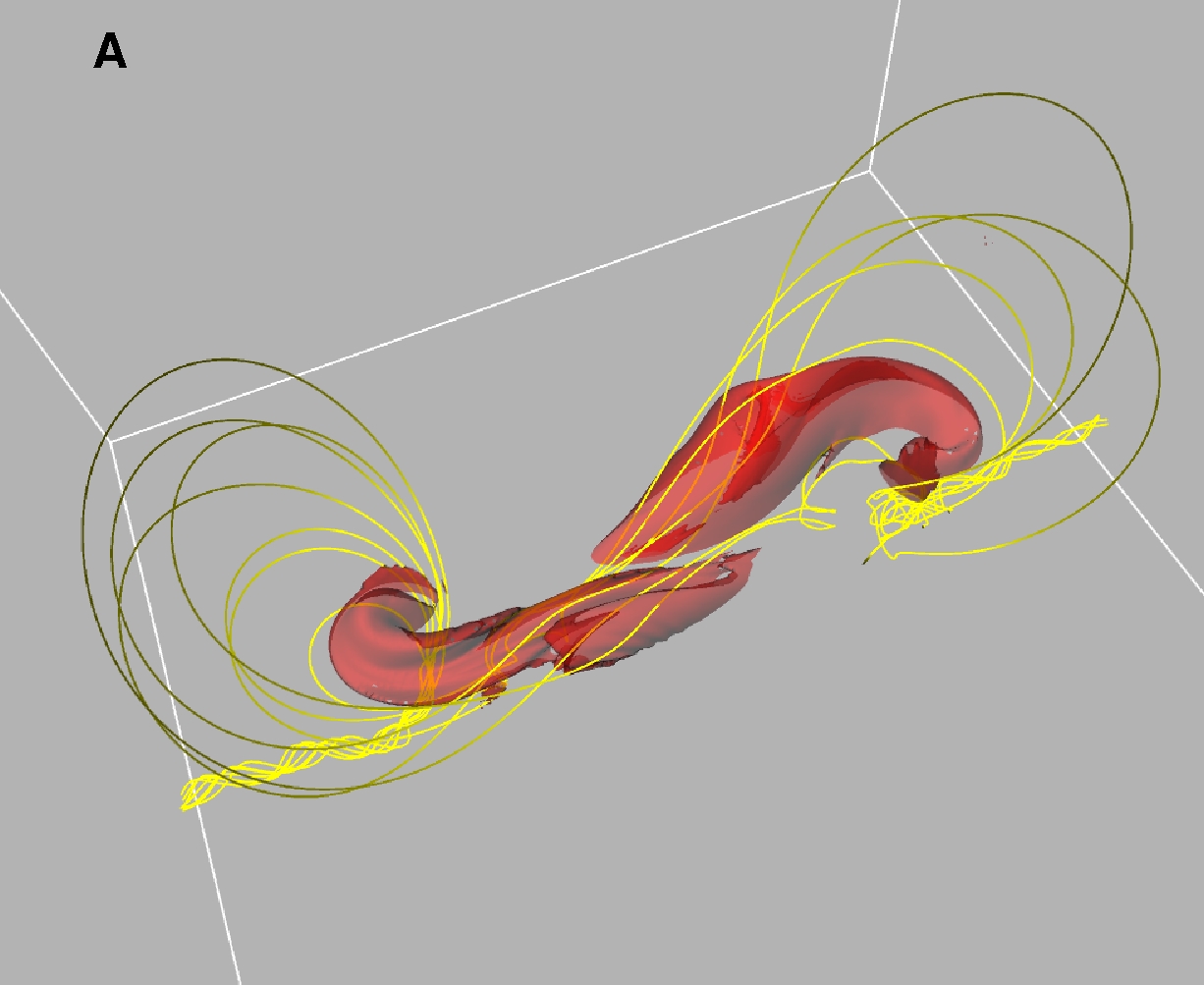}
\includegraphics[scale=0.2]{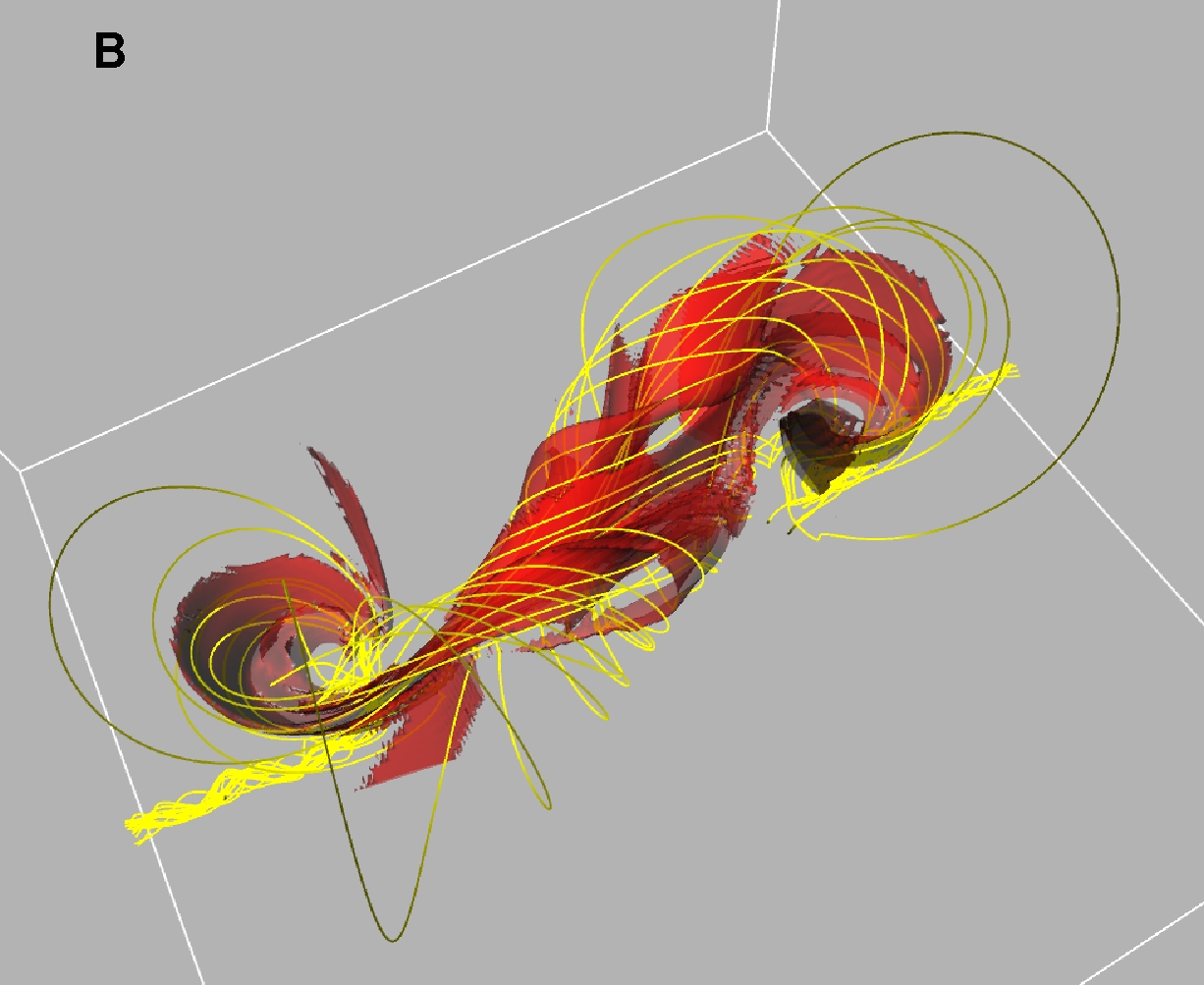}
\caption{The isosurfaces show high current density values, demonstrating the
evolution of a sigmoid. The physical times are (a) $t=50$ minutes and (b) $t=77$ minutes.
The selected field lines are traced from starting points along  the isosurfaces. 
The general evolution of the current density shows that this sigmoid consists of
many current layers (see more in Archontis \textit{et al.} (2009)).}
\label{f9}
\end{figure}

\inlinecite{archontis09c} studied the dynamical evolution 
of sigmoids by examining the three-dimensional structure of current density 
during flux emergence, as illustrated in Figure \ref{f9}. The 
current density (red transparent isosurface) and selected field lines are shown at two 
different times during the evolution. 

Looking at the isosurface in panel A, the current at early times is confined along the two {\it J-like} structures.  
Field lines (in yellow colour) have been traced from the two 
\textit{J-like} segments of the sigmoid. Notice, that \textit{all} the 
field lines  pass below the isosurfaces in going from one end of the sigmoid to the other 
and there are no
field lines that directly connect the positive with the negative
polarity of the emerging field yet.
Eventually, the electric current becomes richer in
structure, as additional current surfaces appear inside  the sigmoidal
volume.

Therefore, at later times (panel B) the current adopts a more pronounced
filamentary structure.  The structure includes a
multiplicity of twisted current filaments and layers, arranged in the
form a sigmoidal shape.  This complexity is apparent not only around
the middle part of the sigmoidal structure but also at the ends. The
latter resemble  spiral scrolls, which in fact become increasingly
twisted during the sigmoid evolution.

\subsubsection{Quasi-Separatrix Layers}
It is worthwhile mentioning that it is likely that the current layers
form along quasi-separatrix layers (QSLs).  In 3D magnetic field
configurations, QSLs are narrow layers where there is a rapid change
in the connectivity of the field lines.  The concept of current sheet
formation and reconnection in QSLs has been studied extensively in the
past few years.
In emergence experiments, a preliminary estimate of the squashing degree
indicates that sites of strong current concentrations develop
preferentially in regions where the squashing degree is large. It is believed 
that QSLs are formed inside the expanding magnetized  volume
of the emerging region due to repeated process of internal
reconnection.

\subsubsection{Heating }
\inlinecite{archontis09c} have further illustrated 
the filamentary structure of the sigmoidal
current and the heating that occurs along it, by producing synthetic
images  of the Joule heating, which is proportional to $|{\bf j}|^2$, and synthetic images of EUV and x-ray intensity
which is related to ${\rho}^2$.  Since extra heating sources (such as heat conduction, radiative transfer, \textit{etc.}) 
were not included in those simulations, this is a very rough approximation to allow comparison with observed quantities. The \textit{heating term} is estimated by

\begin{equation}
\rm{HT} = \int j^2\,dz
\label{Joule}
\end{equation}
and the \textit{intensity term} by
\begin{equation}
\rm{IT} = \int {\rho}^2\,dz.
\label{intensity}
\end{equation}
The line-of-sight integration is performed with respect to height, from the lower photosphere
until well inside corona, and restricting contributions only to plasma with a temperature lying 
between $0.6$ and $2.5$ million K.
\begin{figure}
\centering
\includegraphics[scale=0.2]{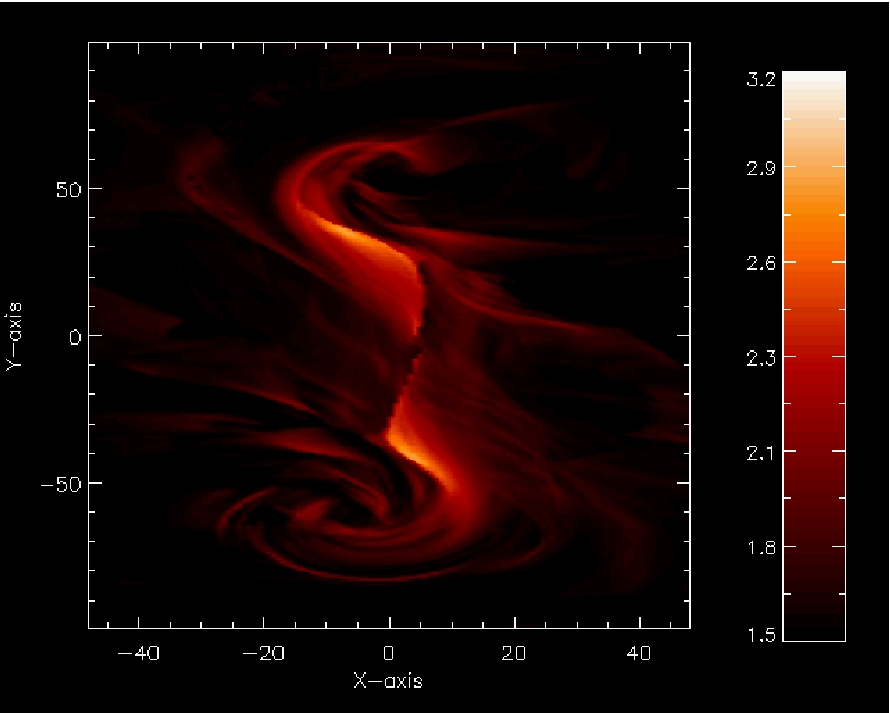}
\includegraphics[scale=0.2]{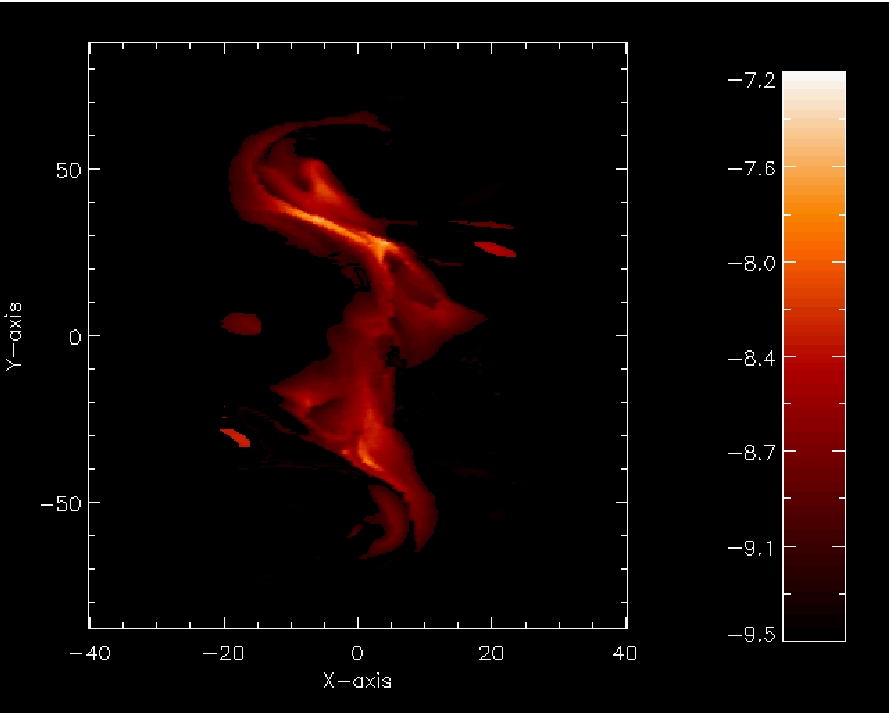}
\includegraphics[scale=0.2]{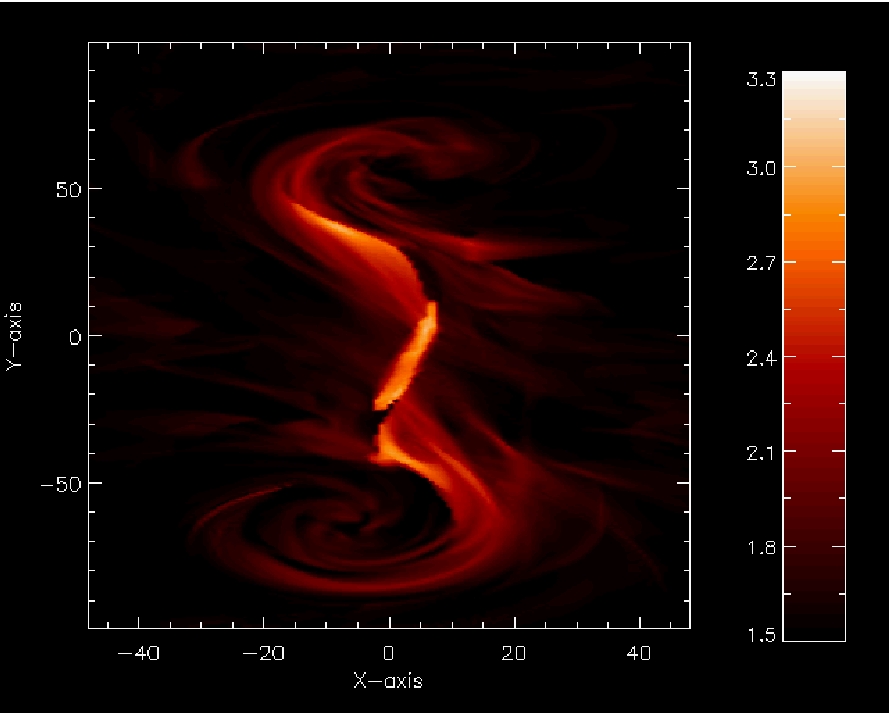}
\includegraphics[scale=0.2]{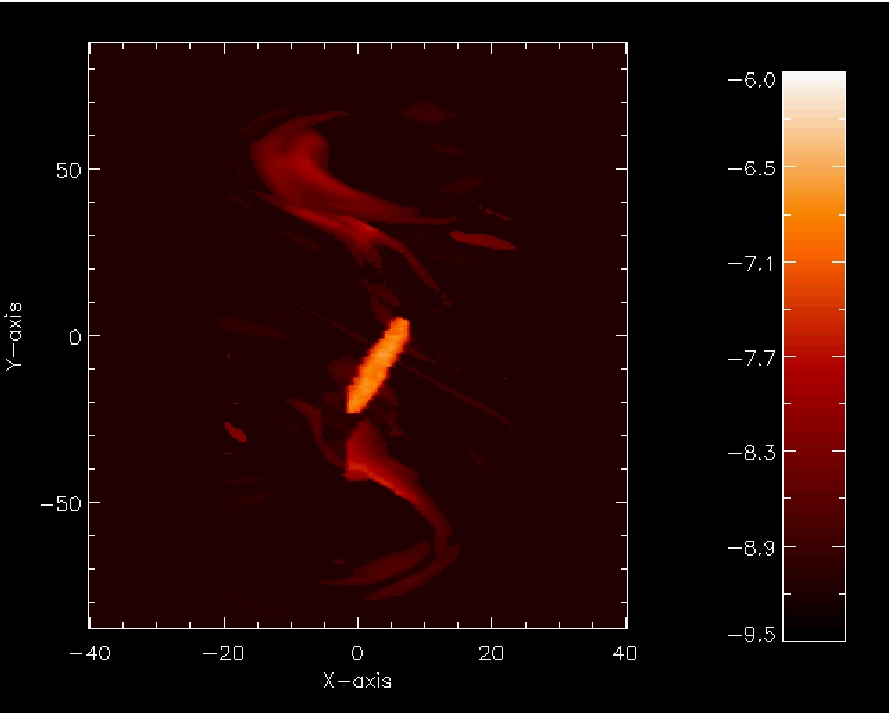}
\caption{The time evolution of the sigmoid is illustrated through the heating term, {HT}, \textit{i.e.} the integration of $|{\bf j}|^2$ 
(left column) and the intensity, {IT}, \textit{i.e.} the integration of ${\rho}^2$ (right
 column) along height. Both terms are shown in a logarithmic
 scale. The times of the two snapshots are $t=70$ minutes (top row)  and $t=77$ minutes (bottom row). Based on Archontis \textit{et al.} (2009).}
\label{f10}
\end{figure}

\subsubsection{Central Brightening and Flaring Activity} 
Figure \ref{f10} shows a visualization of the HT and the IT terms (left and 
right columns respectively), at two different times: 70 and 76 minutes 
(top and bottom row respectively) after the initiation of the emergence. 
At $t=70$ minutes, the plasma, almost along the whole sigmoidal structure, is 
dense. The main heating is initially seen along the elbows, while there is less heating in the central section. 
At the later time (see bottom-left panel), there is a significant amount of
heating at the central part of the sigmoidal structure at the same time as
there is vigorous reconnection of field lines in the narrow
volume between the {\it elbows}. It is interesting that this is the same area
where there was only limited heating at the earlier time. Unlike the earlier evolution, there
is now a current sheet in this central region and the brightening is due to the heating of field lines
by reconnection. A flux tube is formed, as discussed in the above sections, and it rises from this 
central area to the outer  atmosphere. The reconnected field lines are rooted on opposite sides 
of the J-like bundles, mainly at their straight ends in the
central region of the emerging field. The final outcome is
the formation of long, reconnected field lines, that not only connect the two
polarities of the emerging system but also reach up into
the corona.

At lower heights beneath the rising flux tube, the reconnected field lines
create an arcade of short and hot loops. This is very similar to the short \lq
post-flare\rq\ like loops observed in some sigmoids. The simulations of
\inlinecite{archontis08} show similar cusped-like loops.  The bottom-right panel in
Figure \ref{f10} shows that the two {\it elbows}
contain denser plasma than in the surrounding region. The hottest and densest plasma is
of the sigmoidal structure and is spatially coincident with the region of  intense heating in the central
region of the sigmoid.

\subsubsection{Comparison with Observations}
The x-ray Telescope (XRT) on board the \textit{Hinode}
satellite obtained a set of high cadence data, in February 2007, that tracked the formation, evolution and
eruption of a coronal sigmoid. This dataset provided one of the first
observations of such a sigmoidal phenomena with both high spatial and
temporal resolution.  The detailed analysis of the
observations revealed that the overall  S-shape
of the sigmoid is due to many individual loops and is not  defined by
one, single x-ray loop \cite{mcKenzie08}. This is in agreement with the simulations.

\begin{figure}
\centering
\includegraphics[scale=0.4]{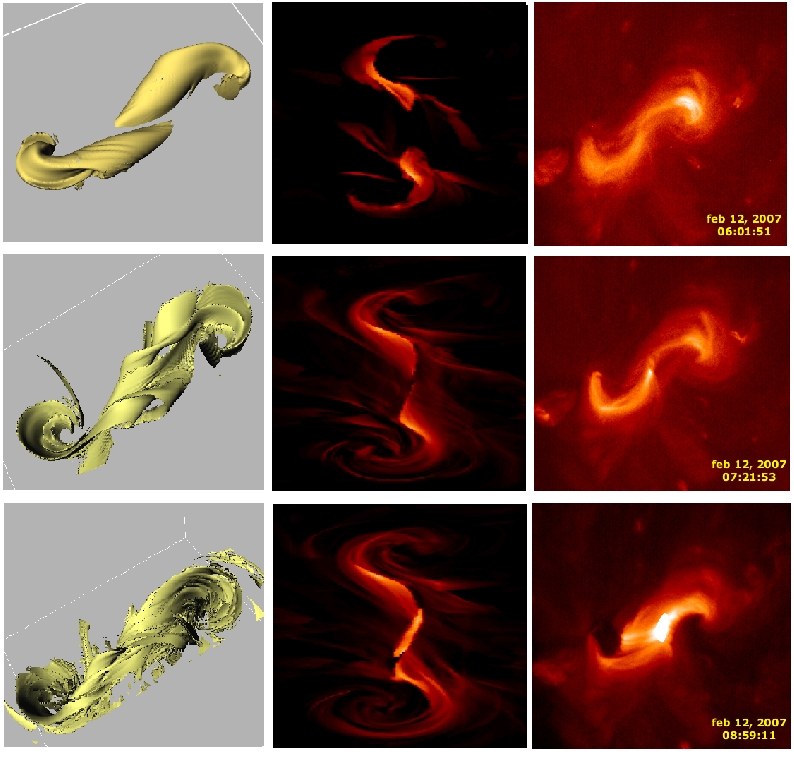}
\caption{ 
Comparison between the numerical experiment (left and middle
columns) and XRT observations  (right column). The evolution of a sigmoid is 
demonstrated through constant current surfaces (left column) and the {\it heating term} (middle column)
XRT images (right column) show the actual sigmoid evolution at three different times. 
From Archontis \textit{et al.} (2009).}
\label{f11}
\end{figure}

A comparison between XRT observations and the results
from the numerical experiments at three different times is presented in Figure \ref{f11}. 
The left column shows isosurfaces of constant current density and the middle column shows
the {\it heating term}, as defined by Equation (\ref{Joule}). 
A series of 
images of the sigmoid from XRT, between 06:01UT and 08:59UT on 12 February, are shown in the right 
column.  A
qualitative comparison between the simulations and observations
identifies some remarkable similarities, both in the geometry and evolution of the
sigmoid. In the top row, two separate {\it elbows} are clearly visible at an early time. 
The S-shaped structure is more confined into the two {\it elbows}, which are dense and hot
as can be seen in the middle top figure.
The subsequent evolution of the structure, prior to the eruption of the ejection of the
new flux rope, is indicated in the middle row. The XRT observations
reveal that  the sigmoid apparently consists of more hot
loops and, from the  simulations, it is likely that these loops are newly
reconnected field lines. The bottom row shows the later stages of the evolution.
XRT observed a brightening at the middle part of the sigmoid and  \inlinecite{mcKenzie08}
suggest that this is due to
a cusped arcade that formed after the eruption of the flux
rope.   \inlinecite{archontis09c} found a
significant rise in temperature and density around
this central part between the two {\it J-like} bundles  of
field lines. The brightening in the
middle only occurred after the ejection of the flux rope from there. Beneath the rising flux tube, 
the reconnected field lines form a hot arcade-like
structure.

\section{Eruptions}\label{SEC:ERUPTIONS}
The previous sections have described some of the main physical mechanisms and properties related to the formation of atmospheric flux ropes, whose axes differ from that of the original emerging tube.  These flux ropes form for a wide choice of field strengths and twists in the initial conditions for the magnetic
field and can be the sources of {\it failed} or {\it successful} eruptions, depending on the presence and form of the overlying coronal magnetic field.  \inlinecite{manchester04} emphasized the importance of the eruptions of flux ropes formed from the self-consistent shear-reconnected arcade.  The new rope consists of field lines with an S-shaped configuration that  may rise with increased speed into the corona.  The early rise of the new
flux rope is driven by pressure forces. Although the eruption of this flux rope has been considered by some as a
CME-like eruption, the initial rise was followed by a deceleration and
then by a phase in which the main forces were balanced.  The maximum speed of the
erupting rope during the simulation is estimated as $32$ km s$^{-1}$.  \inlinecite{manchester04} also reported that the 
field lines of the erupting rope had no valleys and hardly completed a full rotation
in the corona. Therefore, the erupting rope had very little capability for confining a significant
amount of mass.
In addition, the new flux rope, in this model, does not escape but remains trapped by the envelope magnetic field of the original flux tube.  This should be classified as a failed eruption.  

Why does this flux rope result in a failed eruption? When a new flux rope forms through the internal reconnection of emerged
field lines, as discussed in Section \ref{SEC:ROPE}, it is surrounded by un-reconnected, essentially 
north-south field lines (the envelope field) that are
connected to the dense photosphere. These provide a strong anchoring effect through their magnetic tension,
which acts to constrain the flux rope and prevent it from erupting fully. In order to generate a successful eruption, the overlying tension of these un-reconnected field lines must be removed to allow the new flux rope to rise upwards.  One way of achieving this is through reconnection with an overlying coronal field.  This can be an initial equilibrium field \cite{archontis08,dmac09b} or a dynamic one, produced, for example, by the earlier emerging field of another tube \cite{archontis08a}.  Assuming the reconnection between the emerging flux tube and the coronal field is efficient, the tension of the emerging field is weakened and a \lq path is cleared\rq\  for the new flux rope to be ejected upwards. This is similar to the breakout model of \inlinecite{antiochos99} but from a self-consistent evolution.

 \inlinecite{archontis08} study such eruptions with and without an overlying coronal field.  Flux ropes expanding into a field-free corona, in general, produce failed eruptions.  Those ropes that expand into an overlying pre-existing field, however, are capable of producing successful, CME-like eruptions.  They estimate the escape speed to be 200$~$km s$^{-1}$. \inlinecite{dmac09b} perform a similar study but use a toroidal flux tube for the initial condition rather than a cylindrical tube.  They find several eruptions within a single emerging bipolar region.  The first eruption proceeds as in \inlinecite{archontis08}, where internal reconnection within the emerging magnetic field produces a new flux rope and external reconnection between the emerging field and the coronal field allows this rope to escape.  For the first eruption, there is a delay between the formation of the rope and the successful eruption.  Later, when another new flux rope is formed, by the same method as the first, its ejection is almost immediate.  This is because the \lq path\rq\  has been cleared by the eruption of the first rope  and any overlying magnetic tension is much weaker now than it was for the first rope.  Hence, the second rope can escape with greater ease.  Thus, the
existence of an ambient magnetized plasma prior to the emergence is
crucial for the occurrence of fast CME-like eruptions. In contrast to previous simulations, \inlinecite{archontis08} found that successful eruptions bring very dense plasma into the outer solar atmosphere,
due to the supporting effect of U-shaped segments in the field lines of the erupting flux rope. 

\inlinecite{archontis07a} and \inlinecite{archontis08a} reported on the eruption of flux ropes that occurred
in a model using two emerging flux tubes:  the first tube rises to create an
ambient non-uniform coronal field and the second tube emerges into this \lq pre-existing\rq\ 
magnetic field.  The second paper reports on the internal reconnection of
field lines, due to shearing motions  along the polarity inversion line
and horizontal inflows towards the polarity inversion line of the
first emerging flux region. This process is found to be similar to
the tether-cutting mechanism \cite{moore92}, leading to
the  formation of a new flux rope above the original axis of the
emerging system.  It is found that the rising motion of the new
flux rope is affected by the interaction of the two flux tubes. Their
external reconnection releases the downward tension of the ambient
field lines, leading to a faster eruption of the rope.  In fact, it is
the combination of the internal and the external reconnection that
drives the eruptive motion of the plasma. The internal reconnection
forms field lines underneath the rope that exert an upward tension
force to the rope and help it to rise. The external
reconnection removes flux above the rope, opening the way for an
unimpeded eruption.

A different explanation for the formation of the coronal flux rope was
reported by \inlinecite{fan09}. She finds that  substantial rotational motions
of the two polarities of the emerging flux region twist up the inner
field lines  so that they change their orientation into an inverse
configuration. As a result, a flux rope with S-shaped field lines
forms in the corona and rises with increased speed as the rotational
motions continue. The twisting of the flux rope footpoints is
reported as a result of propagation of nonlinear torsional Alfv\'en
waves along the flux tube. Therefore, in this simulation, the
development and rising motion of a coronal flux rope is not due to a
true plasma motion but is due to a propagation effect produced by the
change of horizontal orientations of the emerged field lines. The
vortical  motions at the footpoints of the emerged field lines cause a
Lorentz force that drives the acceleration. However, this
acceleration (similar to the simulations by \inlinecite{manchester04}) does
not seem to lead to a fully dynamical ejection of the  rope into the
outer atmosphere. Within the running time of the simulation, the rope
appears to reach the low corona with a speed of around $20$ km s$^{-1}$.



\section{Conclusions}\label{conclusion}
This paper has reviewed some of the interesting results of flux emergence simulations based
on the resistive MHD equations. Despite the lack of radiative transfer effects and the complex details
of a fully convective solar interior, many significant phenomena and much physical insight can be gained.
The aim of these more idealized models is to correctly follow the magnetic field evolution, as it rises from the
solar interior and emerges into the solar corona. 

The list of physical phenomena that can be compared to observed features is impressive. 
While there are several topics that are not considered in this review, consider, for example, the 
results of the simulations of the emergence and evolution of Active Regions.

\begin{enumerate}
\item Helioseismology can only give a vague impression, at present, of the nature of the magnetic field in
the solar interior. However, it is possible that a detailed parametric study using flux emergence simulations can
provide insight into some of the properties of this interior field, through a comparison
of photospheric phenomena. Studies of the advection of the internal field by
convection motions strongly suggest that the field must be twisted if it is not to be destroyed by the flow. Thus,
the emergence of a twisted flux tube has some specific photospheric characteristics. Photospheric features, such as magnetic tails, provide a strong indication of the amount of twist in the sub-photospheric
flux tube. In addition, the rotation of sunspots can be linked to the untwisting of the sub-photospheric flux tube
through the upwards propagation of torsional Alfv\'en waves. The amount of rotation will be directly related to the amount of twist in the flux tube. Another phenomenon is the sea-serpent nature of
emerging field region. If all the emerging regions are linked to some larger sub-photospheric field, then
the separation between the bipolar regions should be related to the unstable wavelength of the 
Parker instability.
\item The fact that the photosphere and chromosphere are stably stratified means that it is very difficult for the
magnetic field to rise through simple buoyancy. The subsequent emergence of the field requires a magnetic buoyancy instability to occur. The importance of triggering the magnetic buoyancy instability is that
the instability only occurs when the value of $\beta$ at the 
photosphere is of order unity. There are several importance 
consequences of this. Firstly, there is a large amount of magnetic flux left in the interior. Secondly, only 
magnetic fields that are sufficiently strong will emerge properly. Thirdly, since the plasma $\beta$ is of
order unity, the Lorentz force is significant and the magnetic field can cause the plasma to move, rather
than the usual assumption that the magnetic field is moved by the photospheric plasma motions. The
majority of simulations find a shearing motion around the polarity inversion line.
\item All flux emergence simulations show a dramatic adiabatic expansion.  Is there an observed cooling associated with emergence or is there always some heating? A detailed observational study, following
the birth of a new active region right through to its final breakup, is now possible with 
\textit{Hinode} and SDO. The clear link between all the atmospheric levels is now possible.
\item The standard cartoon of flux emergence suggests that the complete flux tube rises into the corona.
However, simulations can easily follow the motion of the flux rope axis. What is clear is that the axis of a
cylindrical flux tube normally only rises to one or two pressure scaleheights into the photosphere. What has been
demonstrated is the importance of the shape of the rising section of the flux tube. When a toroidal shaped
flux tube is used and the field is sufficiently strong, the axis does indeed make it all the
way to the corona. This toroidal shape can be formed either by selecting it at the start or by choosing
an appropriate density deficit profile.
\item A common feature in all simulations is the formation of a new flux rope, where field lines are
weakly twisted around an axis that is not the original flux tube axis. This occurs through a photospheric
shearing motion and a converging flow. This results in reconnection and the formation of a new flux rope.
If the converging flow occurs above the original flux tube axis, then the original axis remains trapped in the 
photosphere and the new rope rises. On the other hand, if the converging flow is beneath the original axis,
then the original axis rises up to the corona. When the shearing and converging flows continue, multiple
flux ropes can form. 
\item Sigmoids naturally appear during flux emergence simulations.
The results from these studies reveal that the dynamical
emergence of a twisted flux tube leads to formation of a sigmoid with
an intricate structure.  Initially the sigmoid comprises of two separate
{\it J-like} features but, as the atmosphere evolves, the internal structure
of the sigmoid becomes more complex.  It now consists of many strong
current layers, which presumably are QSLs, inside which
intensive heating occurs. An interesting issue is whether flux emergence is necessary for the
build-up of complexity in sigmoids.  In our simulations, flux
emergence provides a means to follow the formation and evolution of
sigmoids in a self-consistent manner. It also provides: the
distribution of the line-of-sight magnetic field, $B_z$, at the
photosphere, the photospheric shearing motions and injects the flux tube twist into
the system as helicity.  Thus, it is not unlikely that mechanisms providing
these key ingredients will produce similar sigmoid structures.

\item The formation of new flux ropes, due to reconnection internal to the emerging field, and
the creation of plasmoids, due to reconnection in the current sheet formed
between the leading front of the emerging field and the overlying coronal field both give rise
to dense plasma that is initially ejected upwards. However, the eventual evolution of the
new flux ropes, in particular, depends strongly on whether there is an overlying magnetic field or not. If there 
is no coronal field, then there is no CME-like eruption. If there is an overlying field, then reconnection between
the fields removes the restraining effect of magnetic tension and allows the plasmoid/flux rope 
to rise unhindered.
\end{enumerate}

To date, the idealized simulations have only just begun detailed investigations into the 
physical processes responsible for flux emergence. More work remains to be done. However,
it is essential to bear in mind which specific observational phenomena are being modelled. Do the
general trends seen in the theoretical modelling agree with the typical observations? One major achievement
will be if the results of the simulations can provide information about the nature of the sub-photospheric field and make predictions about observable phenomena arising from the interaction of the emerging and pre-existing 
coronal fields.
 

\begin{acks}
DMacT and VA acknowledge the financial support of the UK Science and Technology Funding Council.  In addition, DMacT acknowledges financial support from the European Commission through the SOLAIRE Network (MTRN-CT-2006-035484). Some of the computer simulations were carried out on the UK MHD Consortium parallel computer at the University of St Andrews, funded jointly by STFC and SRIF.
\end{acks}

\bibliographystyle{spr-mp-sola-cnd} 
\bibliography{references}  
 

	


\end{article} 

\end{document}